\documentclass{aa}
\usepackage{graphicx}
\usepackage{txfonts}
\DeclareMathOperator*{\argmin}{arg\,min}
\DeclareMathOperator*{\argmax}{arg\,max}
\usepackage{algorithm}
\usepackage{algpseudocode}
\usepackage{bbold}
\usepackage{hyperref}

\usepackage{float}
\usepackage[switch]{lineno}
\usepackage{datetime}
\usepackage[usenames,table,dvipsnames]{xcolor}
\usepackage{setspace}
\usepackage[normalem]{ulem}
\usepackage{soul}
\usepackage{subcaption}
\usepackage{dblfloatfix}

\usepackage{makecell}

\definecolor{blue}{RGB}{66, 153, 233}
\definecolor{red}{RGB}{255, 0, 0}
\definecolor{purple}{RGB}{255, 0, 255}
\newcount\Comments  
\newcommand{\kibitz}[2]{\ifnum\Comments=1\textcolor{#1}{#2}\fi}

\begin{document}

   \title{SUSHI: An algorithm for source separation of hyperspectral images with non-stationary spectral variation}

   \subtitle{Semi-blind Unmixing with Sparsity for Hyperspectral Images}

   \author{J. Lascar
          \inst{1}
          \and
          J. Bobin \inst{1}
          \and
          F. Acero \inst{1,2}
          }

   \institute{Université Paris-Saclay, Université Paris Cité, CEA, CNRS, AIM, 91191, Gif-sur-Yvette, France \\
              \email{julia.lascar@cea.fr}
              \and
            FSLAC IRL 2009, CNRS/IAC, La Laguna, Tenerife, Spain \\
             }

   \date{Received July 20, 2023; accepted March 18, 2024}

  \abstract
  {Hyperspectral images are data cubes with two spatial dimensions and a third spectral dimension. They enable the retrieval of a spectrum for each pixel of a camera and thus the mapping of the physical properties of extended sources or collections of point sources.}
  {In this article, we present the Semi-blind Unmixing with Sparsity for Hyperspectral Images (SUSHI), an algorithm for non-stationary unmixing of hyperspectral images with spatial regularization of spectral parameters. The methods allow for the disentangling of physical components without the assumption of a unique spectrum for each component. Thus, unlike most source separation methods used in astrophysics, all physical components obtained by SUSHI vary in spectral shape and in amplitude across the data cube.}
  {Non-stationary source separation is an ill-posed inverse problem that needs to be constrained. We achieve this by 
  training a spectral model and applying a spatial regularization constraint on its parameters.  For the spectral model, we used an Interpolatory Auto-Encoder, a generative model that can be trained with a limited amount of samples. For our spatial regularization, we applied a sparsity constraint on the wavelet transform of the model parameter maps.}
{We applied SUSHI to a toy model meant to resemble the case study of supernova remnants in X-ray astrophysics, though the method may be used on any extended source at any wavelength where hyperspectral images are available. We compared this result to the one obtained by the classic method used in the literature, namely a 1D fit for each individual pixel. We find that SUSHI obtains more accurate results, particularly when it comes to reconstructing physical parameters.
We then applied SUSHI to real X-ray data from the supernova remnant Cassiopeia A and to the Crab Nebula. The results obtained are realistic and in accordance with past findings but have a much better spatial resolution. Thanks to spatial regularization, SUSHI can obtain reliable physical parameters at fine scales that are out of reach for pixel-by-pixel methods. 
} 
{}


   \maketitle
%

\section{Introduction}
\paragraph{}
In ground or space telescopes of various wavelength ranges, many instruments include spectro-imagers.
These tools allow the spectra of many point sources in the field of view to be retrieved at once, and the mapping of extended sources' spectral properties.
Such instruments can obtain what are called hyperspectral images. These cubes of data have two spatial dimensions and a third spectral dimension,
i.e. a spectrum is associated with every pixel of the image.

In extended sources, these spectra will often not come from a single emitting source but from a mix of physical components. For example, the observed spectrum may be the sum of hot gas thermal emission and non-thermal synchrotron emission. In order to analyze these components and extract physical information, it is key to adequately untangle them from the noisy mixed data.
Several approaches can be undertaken to complete this task, which may be categorized into classic pixel-by-pixel fits, stationary unmixing, and non-stationary unmixing methods. Their characteristics are summarized in Table \ref{tab:methods} and detailed in Section \ref{sec:overview}. Non-stationary unmixing aims to account not only for the varying amplitude of each physical component but also for the variation of their spectral shapes across the image.

In this paper, we present the Semi-blind Unmixing with Sparsity for Hyperspectral Images (SUSHI) algorithm, which performs non-stationary unmixing of hyperspectral images using a learned physical spectral model and applying a spatial regularization constraint on the spectral parameters. 
It is applicable to problems where a fixed number of physical components can be assumed and a spectral model for each of those components is available (which need not be analytical).
The method can be used at any wavelength, though here we present its application to the case study of X-ray imagery of supernova remnants, specifically Cassiopeia A, for which we had simulations and real data to test our method on.

In Section \ref{sec:overview}, we present methods existing in the literature to perform the unmixing of hyperspectral images.
In Section \ref{sec:methods}, we first go over the mathematical formalism of the posed problem. Then, we present our trained spectral model, an Interpolatory Auto-Encoder (IAE), before going over our approach of spatial regularization of the spectral parameters.
In Section \ref{sec:results_sim}, we show results on a simulated toy model meant to resemble the supernova remnant Cassiopeia A, and we compare the efficiency of our method to that of the classic approach. Section \ref{sec:results_data} presents the results obtained on real hyperspectral data. Section \ref{sec:results_CasA} presents results on data of Cassiopeia A taken by the Chandra X-ray telescope. Section \ref{sec:crab} shows results obtained on the Crab Nebula, and Sect. \ref{sec:limits} presents an evaluation of current limitations and goes over avenues for future improvements.

\section{Overview of hyperspectral unmixing methods}
\label{sec:overview}
\paragraph{}
In this section, we present three types of approaches used in astrophysics to unmix hyperspectral images: the classic methods, which fit each pixel individually; the stationary unmixing methods, which assumes that there is one spectral shape per component and a varying amplitude; and the non-stationary unmixing methods, for which the spectral shape and the amplitude are allowed to vary. Their characteristics are summarized in Table \ref{tab:methods}.
\subsection{Classic method: Individual pixel spectral fitting}
\paragraph{}
The first approach, which we refer to as the classic method, is to treat each pixel individually. The hyperspectral cube is separated into individual spectra $\textbf{x}_i$ with $n_E$ channels, and a one-dimensional spectral fit is performed on each vector, ignoring the spatial correlation between pixels. The unmixing is simply done by using a model of the following form:
\begin{equation}
    \textbf{x}_i=\sum_{c=0}^{n_C} a^c_i f^c(\theta_i),
\end{equation}
where $n_C$ is the number of physical components; $f^c$ is the model function, which takes in a set of parameters $\theta$ and returns a vector of length $n_E$; and $a^c_i$ is a scalar, the amplitude of component $c$ in the pixel $i$. The unmixing is performed by finding the best-fit parameters $\theta$ and $a$.

This approach is appropriate for pixels that are bright enough, but it is not robust in dim pixels where noise dominates.
The latter case is the most common, and a usual method to alleviate it is to perform a re-binning that combines pixels with low signal-to-noise ratios together. Different methods are used in the 
X-ray astrophysics community, such as weighted Voronoi tessellations \citep{Diehl_2006} or the contour binning method \citep{Sanders_2006}. Re-binning increases the number of counts in each cell but comes at the cost of a loss in spatial resolution and a more complex unmixing task.

To this day, the classic 1D pixel-by-pixel fitting method remains ubiquitous in the X-ray astrophysics community. For instance, it has been applied in  \citet{CHEXMATE_2024}, \citet{Mayer_2023}, and \citet{Sasaki_2022}.

\subsection{Stationary unmixing}
\paragraph{}
A second type of approach involves stationary unmixing or matrix factorization. Such methods consider the hyperspectral image as a 3D matrix $X$ (shape $l\times w \times n_E$) and wish to decompose 
the cube into:
\begin{eqnarray}
    X=\sum_{c=0}^{n_C} \textbf{s}^c  \diamond A^c,
\end{eqnarray}
where for $n_C$ physical components, each has a spectrum $\textbf{s}^c$ (a column vector of length $n_E$) and an amplitude map $A^c$ ($l\times w$) that corresponds to how intense the component is in each pixel. We defined $ \diamond$ as the product that takes in a vector and a matrix to return a 3D tensor: $(\textbf{s} \diamond A)_{i,j,k}= s_k \times A_{i,j}$.
The "stationary" assumption is thus that the spectrum of each component only varies with a factor of amplitude, but the spectral shape is constant over the pixels.

\def\arraystretch{1.4}
\begin{table*}
    \centering
    \begin{tabular}{|m{0.25\textwidth}|m{0.21\textwidth}|m{0.21\textwidth}|m{0.21\textwidth}|}
    \hline
         &\textbf{Classic Method}  &  \textbf{Stationary $\; \; $Unmixing} & \textbf{Non-stationary Unmixing}\\
         \hline \hline
         General Approach & One-dimensional spectral fit of individual pixels. & Matrix factorization. To each component is associated a spectral vector and an amplitude map. & To each component is associated a spectral matrix and an amplitude map. \\
         \hline
         Correlation between pixels & No. Fits pixels individually. & Yes. Treats cube as a whole. & Yes. Treats cube as a whole. \\
         \hline
         Spectral variability & Yes. Varying spectral shapes for each component.  & No, one spectral shape per component.  & Yes. Varying spectral shapes for each component. \\
         \hline
         Examples in Astrophysics & \citet{CHEXMATE_2024},\citet{Mayer_2023},\citet{Sasaki_2022}& GMCA \cite{Adrien_GMCA}, \cite{Bobin_2015} & ROHSA \cite{ROHSA}, SUSHI (this work)\\
         \hline
    \end{tabular}
    \caption{Three types of methods to unmix hyperspectral images with example papers. }
    \label{tab:methods}
\end{table*}

Methods of stationary unmixing include non-negative matrix factorization (NMF), which imposes that all values of $A^c$ are either zero or positive. An extensive review of NMF methods may be found in \cite{review_NMF}. Another type of constraint that can be imposed on $A^c$ (often in addition to an NMF constraint) is that it be sparse under some orthogonal transform, which in turn forms the sparse component analysis (SCA) techniques, as done in \cite{MASS}. If one is justified to assume that the components are statistically independent of each other, the best-suited method is independent component analysis (ICA) as described in \cite{ICA}.

If no prior information is assumed about the spectra $\textbf{s}^c$, the task at hand is one of blind source separation (BSS), as is performed in \cite{Bobin_2015}. When the spectra are given, the source separation is said to be supervised. Methods between those two extremes, such as those that assume that $\textbf{s}^c$ belongs to a certain space of solutions, are referred to as semi-blind, an example of which can be found in \cite{sgmca}.
\subsection{Non-stationary unmixing}
\paragraph{}
While the classic method is hindered by not considering the spatial correlation between pixels, the stationary mixing model makes a non-trivial assumption by stating that each physical component should only have one spectral shape $\textbf{s}^c$. Indeed, in many cases, the spectral shape of a component varies across the image due to physical parameters such as temperature, ionization, chemical composition, velocity effects, among others. In cases where these spectral variations are important enough that the stationary assumption does not hold, a non-stationary model is needed:

\begin{eqnarray}
    X=\sum_{c=0}^{n_C}  A^c\odot S^c ,
\end{eqnarray}
where $S$ is now a tensor of size ($l\times w \times n_E$), the same shape as X, and we define $\odot$ as the product that multiplies a 3D tensor and a 2D matrix as follows: $(A\odot S)_{i,j,k}= A_{i,j}\times S_{i,j,k}$.
Now, for each component $c$, the spectral shapes can vary across the image in addition to their amplitude.

In the remote sensing community, many methods have been studied to account for spectral variations of the unmixed components \citep[for a comprehensive review, see][]{Borsoi_2021}, but these methods make several assumptions that make them inadequate for astrophysical data. They usually assume (low) additive Gaussian noise, the so-called pure-pixel assumption (where for each component, there is at least one pixel that only contains that component, or approximately so), and normalized abundances. In the case study of X-ray astrophysics, as in many other astrophysical data, these assumptions rarely hold.

Further, in order to obtain scientifically interpretable results in the context of astrophysics, it is essential to fit a physical model for each component. This is necessary if we want to map physical parameters across a source, which is a key objective when studying astrophysical objects. In the remote sensing literature, including a varying physical model is usually done to account for the scattering of light off of surfaces.
In particular, this was done in the seminal work of \cite{Hapke}, who provided an approximate solution to the radiative transfer equation. The key difficulty with including the Hapke model is that it requires knowledge of many parameters about the viewing geometry of the scene and scattering properties of the materials in order to be mathematically tractable, properties which are often unknown. Remote sensing methods have thus focused on simplifying the assumptions of the Hapke model, for instance, to retrieve the extended linear mixing model \cite{drumetz2019spectral}. Alternatively, works such as \cite{Shi2022} have opted to forgo a physics-based model and instead train a model on the physical variations from the data itself.

In astrophysics, often the interest is in unmixing components whose spectra are all assumed to follow a known model, but non-stationary unmixing methods have rarely been proposed. One approach was developed by \cite{ROHSA}, with the ROHSA algorithm, a Gaussian decomposition algorithm that uses a Laplacian filter on the Gaussian parameters to unmix hyperspectral images. While this method works very well for the case study of gas phase separation of the hydrogen line emission at 21 cm in radio data, and any other data where Gaussian decomposition is relevant, it will not be applicable to cases where a Gaussian model is not advised.

SUSHI is a non-stationary unmixing method that uses a learned model as a surrogate and applies a constraint of spatial regularization on the model's parameters. In the next section, we explain what that entails and detail our methodology.--------------------------------------------------------------------
\section{Methodology}
\label{sec:methods}
\subsection{Mathematical context}
\paragraph{}
Let the data we wish to analyze be a noisy hyperspectral cube $X$ with $n_P=l\times w$ pixels and $n_E$ energy channels. The observed data consists of $n_C$ entangled physical components, and the objective is to unmix them in order to obtain one cube per component so that
the result is closest to the underlying physical truth. If $\tilde{X}$ is the ground truth, our non-stationary mixing model is
\begin{equation}
    \tilde{X}= \sum_{c=0}^{N_{C}}\tilde{A}^c \odot \tilde{S}^c,
    \label{eq:mixingmodel}
\end{equation}
where for every component $c$, $\tilde{A}^c \in \mathbb{R}^{n_P}_+$ is the amplitude map, 
which is a matrix with a scalar value for each pixel corresponding to the brightness of the component in that pixel; $\tilde{S}^c \in \mathbb{R}^{n_P\times n_E}$
is the spectral shape matrix, which for each pixel contains a normalized spectrum; and $\tilde{A}$ captures the intensity of each component in different areas of the image, whereas $\tilde{S}$ captures the spectral variations across the image.
\\\\
Our objective was, given a noisy $X$, to find a solution $\{\hat{A},\hat{S}\}$ closest to this ground truth. We tackled this problem via a likelihood maximization scheme:

\begin{equation}
    \{\hat{A},\hat{S}\} = \argmax_{A,S} \mathcal{L}( X \,|\, A, S),
    \label{eq:genproblem}
\end{equation}
where $\mathcal{L}$ is the likelihood to maximize.
In our test case of X-ray astronomy, where the properties of each photon can be measured individually (count statistics domain), we used the Poisson likelihood, and the problem can be reduced to minimizing the negative log-likelihood:
\begin{equation}
    \{\hat{A},\hat{S}\}= \argmin_{A,S} \sum_{c=1}^{n_{C}}
    \sum_{i,j=1}^{l,w}
    \sum_{k=1}^{n_E}
    A^c_{(i,j)} S^c_{(i,j,k)}-ln\Big(A^c_{(i,j)}S^c_{(i,j,k)}\Big)X_{(i,j,k)},
    \label{eq:negloglikelihood}
\end{equation}
though this likelihood could be replaced by an $\chi^2$ or a custom likelihood function if needed.

The problem posed in Equations \ref{eq:genproblem} and \ref{eq:negloglikelihood} is an ill-posed inverse problem in the Hadamard sense \citep{Hadamard}, meaning that it allows for more than one solution and that the solution does not necessarily vary smoothly with changes in the data $X$. 
Inverse problems involve a process of inferring the causal factors behind a set of observations, and their ill-posedness arises when not enough information is taken into consideration to solve the problem. This occurs, for instance, when a linear system of equations has fewer equations than unknowns.
\\\\
The ill-posedness of a problem can be resolved by constraining the space of admissible solutions. This can be done by picking a model to fit the data and by regularizing the problem, that is, constraining it, such as by imposing boundary conditions or introducing another competing term to our cost function. The choice of model and of regularization are key. If they are poorly chosen, they will introduce a bias to the solution.
\\\\
For the model, in our situation of hyperspectral unmixing, the spectrum of each component is assumed to follow a physical model $\mathcal{M}$ with parameters $\theta$, and thus $S^c=\mathcal{M}^c({\theta}^c)$. This will ensure that our solution is scientifically interpretable rather than simply a close fit to the data set. However, this model is rarely analytical, and in some cases would need to be convolved with the instrument response functions so it may not be easily accessible. It may be costly to compute and/or non-differentiable. These two properties are key, seeing as accounting for both a spectral model and spatial regularization requires iterative numerical solvers that can be used only for differentiable models, such as \cite{PALM}.

To mitigate these limitations, we chose to use a learned representation of the spectral model, which is detailed in Section \ref{sec:IAE}.

For the spatial regularization, we took advantage of the fact that the spectral parameters are likely to vary smoothly across the image, which is mostly the case in astrophysical extended sources. Neighboring pixels are unlikely to display drastic differences. Thus, a spatial regularization on the spectral parameters is a good choice, and it is detailed in Section \ref{sec:spatreg}. This approach offers the advantage that even if some pixels are dominated by noise, the information brought by neighboring pixels will help with the unmixing process.

Thus, the problem stated in equation \ref{eq:genproblem} becomes
\begin{equation}
    \{\hat{A},\hat{\theta}\} = \argmax_{A,\theta}  \mathcal{L}\Big(X | A,\theta\Big)-\rho \sum_c^{n_C} \varphi(\theta^c),
    \label{eq:fullproblem}
\end{equation}
where $\varphi$ is our spatial regularization function and $\rho$ is the regularization parameter that fine-tunes the balance between
the spatial regularization term and the data-fidelity term.The two following sections explain our chosen spectral model and spatial regularization schemes in more detail.
\subsection{Learned spectral model}
\label{sec:IAE}
\paragraph{}
In cases where an analytical model $\mathcal{M}$ is available for every physical component, one simply fits the data to the model, and well-known methods of optimization can be applied. But often an analytical model is not available. For instance, $\mathcal{M}$ might require costly Monte Carlo simulations, the use of look-up tables, or convolution with complex instrument-response functions.
\\\\
In those examples, the model will be computationally expensive; further, it will not be differentiable, which is a hindrance for optimization, as minimization methods typically use the gradient of the cost function to find the optimal solution. There are minimization methods that approximate the gradient, such as the Nelder-Mead method, but to use state-of-the-art methods for regularized optimization problems (such as proximal gradient descent methods), we needed a model $\mathcal{M}$ that is exactly differentiable. 
Thus, we sought to train a surrogate model $\mathcal{M}^{\star}(\theta)$ that would have the following characteristics: \textbf{(1)} $\mathcal{M}^{\star}$ is not costly to call; \textbf{(2)}$\mathcal{M}^{\star}$ is differentiable so as to be plugged in iterative solvers; and \textbf{(3)} the model parameters $\theta$ vary smoothly with changes in physical parameters (so we could apply spatial regularization on $\theta$ maps).

\begin{figure}
    \centering
    \includegraphics[width=\linewidth]{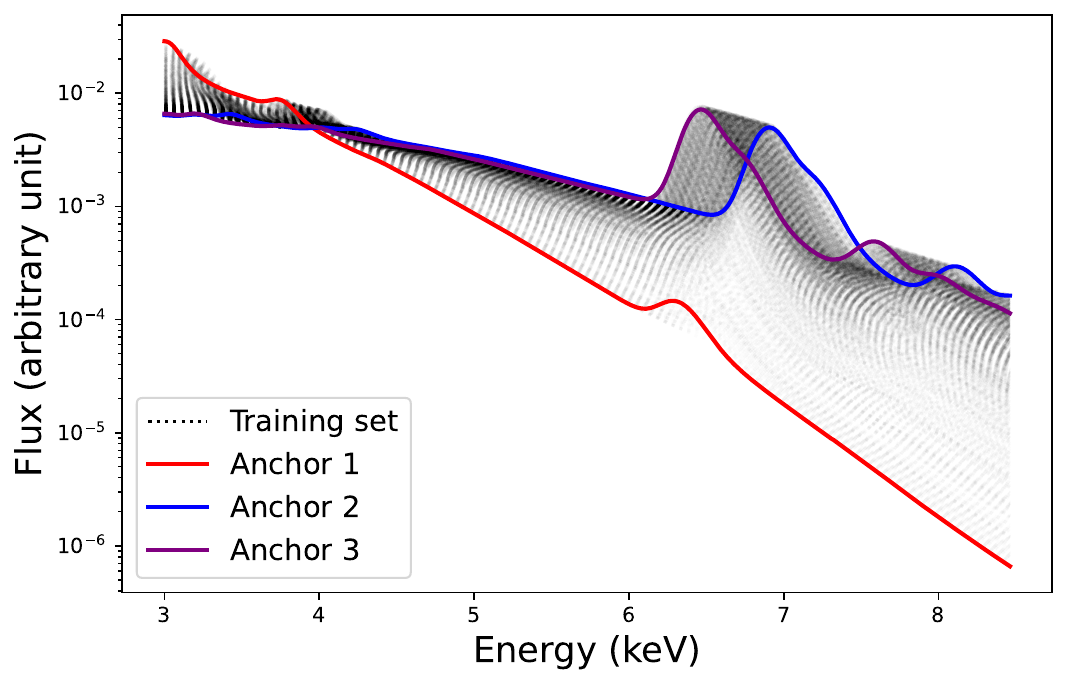}
    \caption{Training set for the IAE model for the emission spectra from collisionally ionized hot gas around the Fe K line. There are two underlying physical parameters: temperature, varying between [0.8 keV, 6 keV], and redshift, varying between [-0.033,+0.033]. Anchor points are in color: Red is for the lowest temperature, lowest redshift. Purple shows the highest temperature, lowest redshift. Blue is for the highest temperature, highest redshift.}
    \label{fig:IAE_trainingset}
\end{figure}

For this purpose, we used a model learned by a neural network called an Interpolatory Auto-Encoder (IAE). The principle behind the IAE is briefly described in this section, though the interested reader can find a more detailed presentation in \cite{IAE} and \cite{sgmca}. The IAE's purpose is to learn a physical model (here, a 1D spectral model) from a limited amount of training samples. It does so by learning to interpolate between well-chosen examples called anchor points.
\\
\\
Figure \ref{fig:IAE_trainingset} shows the spectral training set for the IAE trained for our case study of X-ray astrophysics. The example we considered is an emission spectrum from collisionally ionized hot gas (APEC model\footnote{\url{http://www.atomdb.org/physics.php}}) convolved with the Chandra X-ray telescope instrumental response. In the figure, the $N_A$ anchor points $\mathcal{A}$ are in color, while the black curves are examples of what spectra interpolated between those anchor points would look like (i.e., examples of what the IAE should learn to generate). The physical model used to generate this training set is accurate for our purpose, but because it is based on look-up tables, it has the aforementioned problems of being costly to call and non-differentiable.

\begin{figure}[h]
    \centering
    \includegraphics[width=\linewidth]{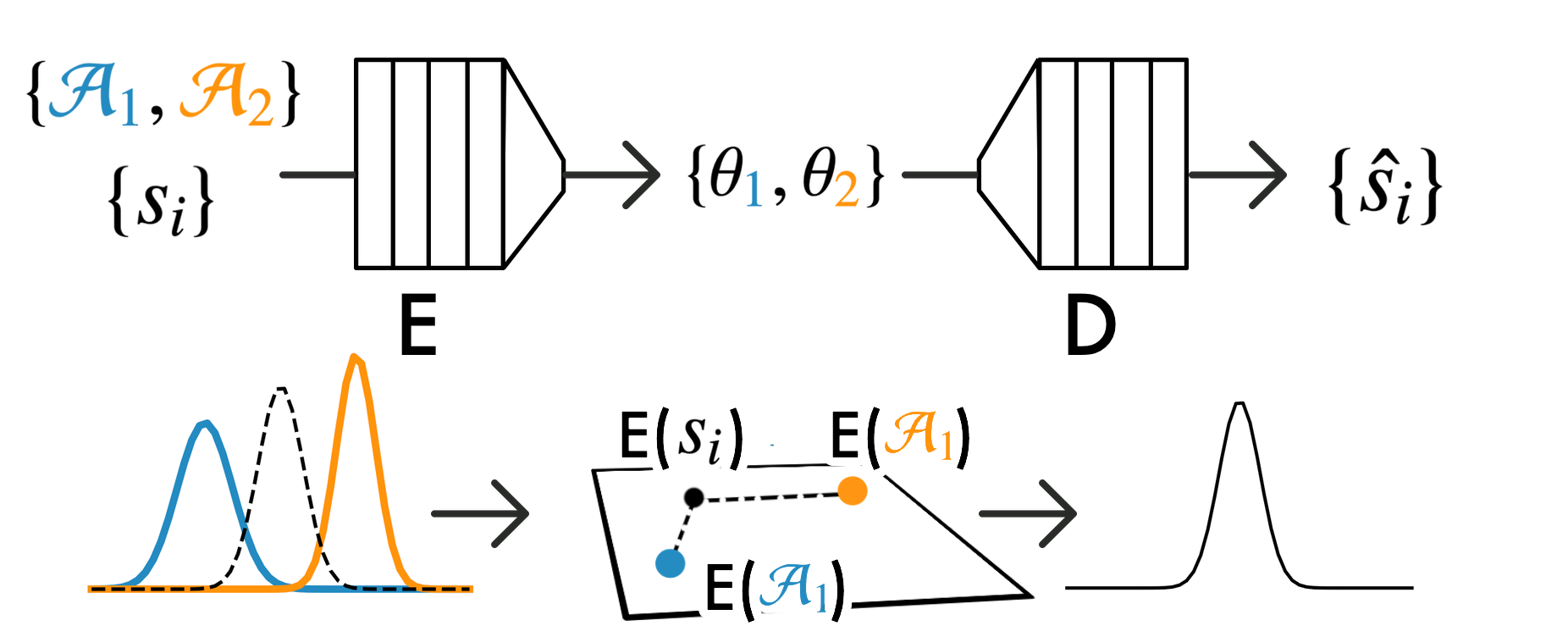}
    \caption{Diagram of the IAE architecture for the simple case with $n_A=2$ anchor points $\mathcal{A}_1$ and $\mathcal{A}_2$, shown in blue and orange. In black, $s_i$ is an example member of the training set. The encoder \textbf{E} learns how to reduce the dimensions of $s_i$ ($n_E$ energy channels) into two parameters, $\{\theta_1,\theta_2\}$, such that in the latent space, these correspond to the weight of each encoded anchor point for $\textbf{E}(s_i)$ in order to be linearly interpolated between them. The decoder $\textbf{D}$ learns how to return $\hat{s_i}$ close to the input $s_i$from $\{\theta_1,\theta_2\}$.}
    \label{fig:IAE_architecture}
\end{figure}

\begin{figure}
    \centering
    \includegraphics[width=\linewidth]{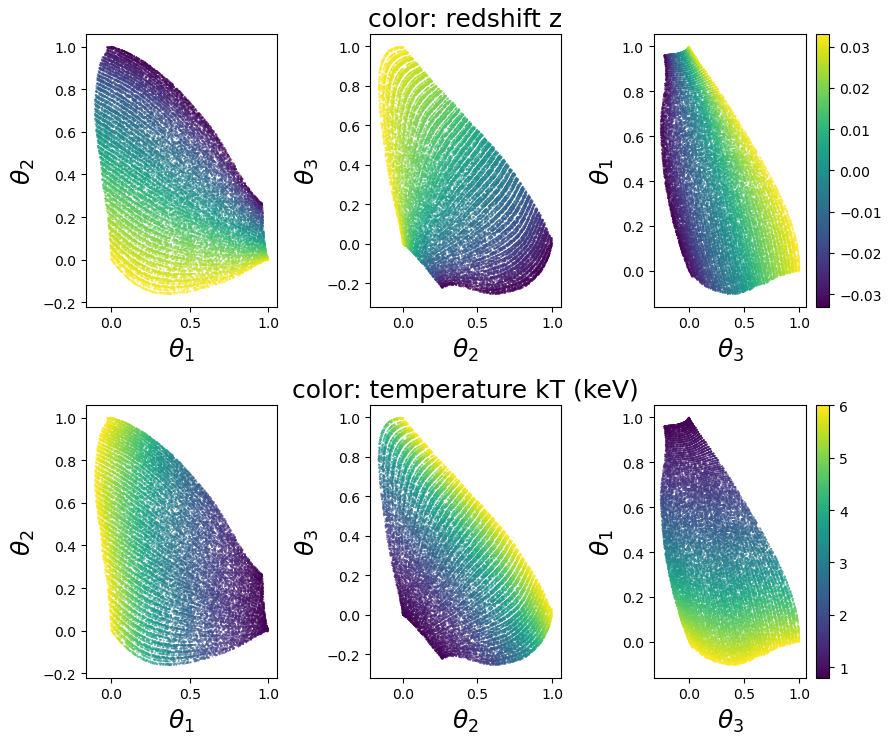}
    \caption{Distribution of the three training set's latent parameters $(\theta_1,\theta_2,\theta_3)$ for the IAE trained on the hot plasma X-ray emission spectra with two variable physical parameters. The color bars show the relationship with plasma temperature kT and velocity redshift z.}
    \label{fig:lambda_distrib}
\end{figure}

Inspired by the architecture of auto-encoders, the IAE learns an encoder \textbf{E} and a decoder \textbf{D} function that can respectively transform a signal toward and back from a latent manifold. Both of these functions are multilayer perceptron with $L$ layers and parameters $\sigma$ (though \textbf{E} and \textbf{D} have distinct weights). On the latent manifold, the interpolation between anchor points becomes a linear interpolation (whereas it would be non-linear in the physical space). For a training set $\mathcal{T}$, the cost function $C$ to minimize in order to learn the parameters of  \textbf{E} and \textbf{D} is

\begin{equation}
\label{eq:costfuncIAE}
    C(\sigma)= \sum_{s_i\in \mathcal{T}}\big|\big| \textbf{E}_{\sigma}(s_i)-(I\circ \textbf{E}_{\sigma})(s_i))\big|\big| _2^2+ \mu \sum_{s_i\in \mathcal{T}}\big|\big|
    s_i-(\textbf{D}_{\sigma}\circ I\circ \textbf{E}_{\sigma})(s_i)
    \big|\big|_2^2,
\end{equation}
where $\mu$ is simply a term to control the relative impact of the two terms, $s_i$ is an element from the training set, and $I$ is a linear interpolator function, which in the latent space projects the signal in the anchor points' barycentric span, that is, $(I\circ \textbf{E})(s_i)=\sum_a^{N_A} \theta_a \textbf{E} (\mathcal{A}_a).$
Here, $\theta_a$ is the weight of each anchor point $\mathcal{A}_a$ for the point projected from $\textbf{E}(s_i)$.
\\\\
Thus, the first term in Equation \ref{eq:costfuncIAE} ensures that \textbf{E} will map to the space of spectra interpolated between the anchor points. The second term ensures that \textbf{D} will return a spectra close to those shown in the training set. A simplified diagram of the IAE architecture is shown in Figure \ref{fig:IAE_architecture}.
\\\\
The function of particular interest to us is the decoder \textbf{D}. Once learned, it takes as input a set of $N_A$ weights and returns a spectrum in the space interpolated between the anchor points, which will be a physically interpretable spectrum. Thus, $\textbf{D}(\theta)$ fits the requirement for our model $\mathcal{M}(\theta)$ set out at the beginning of this section:
\textbf{(1)} Once trained, \textbf{D} is fast to call upon. For our case study of X-ray astrophysics, a call to \textbf{D} was over 60 times faster than calling the physical model based on look-up tables (XSpec model in Sherpa\footnote{https://cxc.cfa.harvard.edu/sherpa/} ).    \textbf{(2)}\textbf{D} is exactly differentiable, and its derivative can be evaluated using auto-differentiation since it is a standard multilayer perceptron.
\textbf{(3)} $\theta$ parameters vary smoothly with physical parameters thanks to the interpolarotary structure. This can be seen in Figure \ref{fig:lambda_distrib} for the thermal component in our case study.

An IAE model may be trained for each physical component. The code to do so is openly accessible on GitHub\footnote{\url{https://github.com/jbobin/IAE}}. Details regarding the specific architecture of the IAE models used in this work and evaluation of their quality are available in the Appendix, Section \ref{sec:Appendix_IAE}.
\subsection{Spatial regularization: Sparsity constraint}
\label{sec:spatreg}
\paragraph{}
If we treated each spectrum individually, we would not be taking full advantage of the hyperspectral data's structure. For most extended sources, be they supernova remnants, diffuse gas in galaxies, or galaxy clusters, neighboring regions in the image will be in direct physical interaction with one another, and thus, physical parameters will vary smoothly from one pixel to the next, provided the spatial resolution is not too coarse. This implies that the spectral shapes will also vary smoothly.
Hence, by picking a well-chosen spatial regularization, we could better separate the sources and obtain a more accurate spectral fit in pixels with a low signal-to-noise ratio and/or when one physical component drowns out the others.

A common and efficient choice of smoothness regularization is a sparsity constraint in an adapted signal representation, such as the Fourier transform or wavelet transforms. That is to say, if we apply a well-chosen transform on the data, it will contain mostly zero (or almost zero) coefficients $\mathcal{C}$. A threshold can then be picked, and all coefficients under that threshold can be put to zero,
as they are assumed to be caused by noise. When the inverse transform is applied to these coefficients, the data is smoother and displays less noise.

Promoting the sparsity of the coefficients $\mathcal{C}$ can be done by minimizing their $l_1$ norm, defined as $||x||_1=\sum_i|\mathcal{C}_i|$ (see \cite{sparsity}).
 Thus, the equation \ref{eq:fullproblem} becomes
\begin{equation}
    \{\hat{A},\hat{\theta}\} = \argmax_{A,\theta}  \mathcal{L}\Big(X | A ,\theta\Big)
    - \rho \sum_c^{n_C} ||\mathcal{C}(\theta^c)||_1.
    \label{eq:fullproblem_regl1}
\end{equation}

For images such as our maps of spectral parameters $\theta^c$, wavelet transforms — which capture multiscale information — are especially appropriate. In this work in particular, we used the isotropic undecimated wavelet transform introduced in \cite{starlet}, dubbed "starlet transform". This transform is especially useful for astrophysical data given that such images often look isotropic, but for other applications, it may be replaced by other transforms, such as a curvelet transform (for filamentary structures).

A more detailed presentation of the starlet transform may be found in \cite{starckbook}. For our purpose, it is important to know that it decomposes an image $X$ of size $n\times n$ into a coarse scale $w_c$ (the smoothest scale) and fine scales $\mathcal{W}=\{w_1,...w_J\}$, capturing the details from the finest to the smoothest. Each of these scales has a size of $n\times n$, and $J$ is the chosen number of scales.

By using our example of Poisson statistics and plugging in the negative Poisson log-likelihood defined in Equation \ref{eq:negloglikelihood} (though again, this could be replaced by another likelihood) and the starlet transform fine scale coefficients $\mathcal{W}$, we thus obtained the following cost function to minimize:
\begin{align}
    \{\hat{A},\hat{\theta}\} = \argmin_{A,\theta}& \sum_{c, i,j, k=1}^{n_{C}, l, w, n_E}  \Bigg[
    A^c_{(i,j)} \mathcal{M}^c({\theta}^c)_{(i,j,k)} \nonumber
    \\
    &-ln\Big(A^c_{(i,j)} \mathcal{M}^c({\theta}^c)_{(i,j,k)}\Big)X_{(i,j,k)} \Bigg] + \rho ||\mathcal{W}(\theta^c)||_1,
    \label{eq:fullproblem_regl1_poisson}
\end{align}
with every $A_{(i,j)}^c$ greater than or equal to zero.

The equation is thus aimed at achieving two competing tasks: ensuring data fidelity by maximizing the likelihood (likelihood term) and making the spatial variations of the spectral parameters smooth by ensuring the wavelet coefficients of the $\theta$ maps are as sparse as possible (regularization term). If the first term is prioritized and the second term is neglected, there will be overfitting — the noise will induce artifacts in the fit. But if the second term is prioritized too much, the fit will be too smooth and lead to a biased result. Thus, the result will be a trade-off between data fidelity and regularization, the relative importance of which is controlled by the regularization parameter $\rho$.

\subsection{Algorithm description}
\paragraph{ }

The problem posed in Equation \ref{eq:fullproblem_regl1_poisson} is not convex as a whole, but multiconvex. That is to say, it is convex with respect to blocks of variables. In such a framework, a natural choice of optimization algorithm is the proximal alternating linearized minimization (PALM) algorithm \cite{PALM}, which consists of updating each block while keeping the others fixed, alternating between gradient descent steps (to promote data fidelity) and proximal steps (for regularization).

In our case, the model $\mathcal{M}(\theta)$ is not convex, but in practice it is locally convex if the IAE is well trained \citep[see ][ Section 2.4.2 for more detail]{remi_thesis}, so we used a PALM architecture.

Thus, for each component, the algorithm consists of three main updates (while keeping all else fixed):
First, a gradient descent step to minimize the negative likelihood with respect to $\theta^c$. Then, a proximal step to minimize the norm of $\mathcal{W}(\theta^c)$ while remaining close to the previously found value of $\theta^c$.
Another gradient descent step follows, now to minimize the negative likelihood with respect to $A^c$. Finally, a proximal step to ensure $A^c\geq 0$.

This should be done for each component, then iterated until a stopping criterion is reached. The pseudo-code can be found in Algorithm \ref{alg:sushi}. The following subsections detail each step of the algorithm more explicitly.

\subsubsection{Initialization}

As input, the algorithm takes a data cube $X$, trained IAE models (see Section \ref{sec:IAE}) $\{\mathcal{M}^0, ..., \mathcal{M}^{n_C}\}$, the cost function $\mathcal{L}$ (for instance, the negative log-likelihood), and hyperparameters that control the intensity of the spatial regularization, 
namely, the number of wavelet scales $J$ and the sparsity threshold $k$ (a factor of the regularization parameter $\rho$; explained further in Section \ref{sec:regalgostep}).
For the first guess in amplitude, we chose the same for every component, assuming an equal amplitude on all pixels, corresponding to a fraction of the sum of counts over the spectra for that pixel: $A_{0,(i,j)}^c=\sum_k^{n_E}X_{(i,j,k)}/n_{C}$. For our first guess $\theta^c$, we took the same value for all pixels such that the sum of the vector $\theta^c_{i}$ is equal to 1 $\theta^c_{i}=[1/n_A^c,...,1/n_A^c]$.

\subsubsection{Gradient descent step over $\theta^c$}
Gradient descent is a common way of finding the parameters that minimize a function and is based on iteratively subtracting those parameters with the function's gradient. In this case, we first performed a gradient descent step on the $\theta^c$ parameters to go toward the parameters that best minimize the likelihood while keeping all other components and variables fixed. For our case study where $\mathcal{L}$ is the Poisson negative log-likelihood, the cost function is the first part of Equation \ref{eq:fullproblem_regl1_poisson}:
\begin{align}
    \mathcal{L}(A,\theta)= \sum_{c, i, j, k=0}^{n_{C}, l, w, n_E}  \Bigg[
    A^c_{(i,j)} \odot \mathcal{M}^c({\theta}^c)_{(i,j,k)} 
    &-ln\Big(A^c_{i} \odot \mathcal{M}^c({\theta}^c)_{(i,j,k)}\Big)X_{(i,k)} \Bigg]
    \label{eq:cost_function}.
\end{align}

Its partial derivative is
\begin{eqnarray}
    \frac{\partial \mathcal{L}}{\partial \theta^c}=A\odot \frac{\partial \mathcal{M}(\theta^c)}{\partial \theta^c}^T \Bigg( \mathbb{1}_{m\times n \times l} - \frac{X}{\mathcal{M}}\Bigg),
\end{eqnarray}
which is not explicitly analytical because $\frac{\partial \mathcal{M}}{\partial \theta^c}$ is not, but since $\mathcal{M}$ is a standard multilayer perceptron neural network, it can be easily calculated with an auto-differentiation scheme.
\paragraph{ }
When it comes to the gradient descent step size $\alpha_{\theta^c}$, an ideal choice is to pick the inverse of the second derivative (taking the Hessian diagonal elements). The second derivative of $\mathcal{L}$ over $\theta^c$ could also be calculated via auto-differentiation, but it adds an unnecessary complexity that slows down the algorithm more than it speeds up convergence. Since the gradient scales to the amplitude, as an approximation, we used
\begin{equation}
\alpha_{\theta}^c=\frac{1}{10\max(A_0)^c},
    \label{eq:stepsizetheta}
\end{equation}

where $A_0$ is the amplitude at iteration $t=0$.
Since all the latent $\theta$ parameters naturally remain in values within the same order of magnitude ($\sim$1), the step size should not vary much and can be kept constant for all pixels.

\subsubsection{Regularization step for latent parameters}
\label{sec:regalgostep}
After the gradient descent step over $\theta^c$ comes the step that regularizes $\theta^c$ maps. The idea is to apply a proximal operator, \textbf{prox} \citep{prox}. When trying to minimize a regularizing function $\varphi$ while keeping close to the value $x$ obtained by minimizing a function $\mathcal{L}$, the proximal operator $\textbf{prox}_f(\varphi)$ returns the value $y$ that minimizes $\varphi(y)$ while keeping as close as possible to $x$. Formally,
\begin{equation}
    \textbf{prox}_{\varphi}(x)=\argmin_y(\varphi(y)+\frac{1}{2}||y-x||_2^2).
\end{equation}
For the $l_1$ loss function, the proximal operator is a soft thresholding function:
\begin{equation}
\label{eq:proxl1}
   \textbf{prox}_{l_1,\rho}(x)=
    \begin{cases}
     0 & \text{if } |x| < \rho\\
     x - \rho . sign(x) & \text{if } |x| \geq \rho \\
    \end{cases},
\end{equation}
where $\rho$ is the regularization parameter, the same as in equation \ref{eq:fullproblem_regl1}, multiplied by the gradient descent step. Following \cite{mad}, we used the median absolute deviation $\textbf{mad}(X)=1.4826\times median(|X-median(X)|)$ (a robust measure of variability) to define $\rho$.
In the case of normally distributed data, $\textbf{mad}$ is a consistent estimator of the standard deviation $\sigma$. Indeed, $\hat{\sigma}(x)=median(|x-median(x)|)/(\Phi^{-1}(3/4))\approx 1.4826\times median(|x-median(x)|)$, where $\Phi^{-1}$ is the reciprocal of the quantile function for the normal distribution \citep{mad}.

In our case, it would be unjustified to claim that the wavelet coefficients of the spectral parameters follow a normal distribution. The threshold should thus be fine-tuned empirically if needed, but choosing the standard $\rho=k\times 1.4826$  provided satisfactory results. The k factor we chose was one, but this choice should be made according to the data. If the data has great spatial resolution and is expected to be very sparse, a high k (such as k=3) will reconstruct the ground truth with great accuracy. In cases when the data is not expected to be so sparse, choosing a high k will overly smooth the data (some of the signal's coefficients will be under the threshold). A lower k (k=1 or k=2) would then be more adequate, as it would not erase low coefficients.

\subsubsection{Gradient descent step over the amplitude}

This step is similar to the gradient step over $\theta^c$, but in this case all parameters are kept constant except the amplitude of the component $c$. The gradient is calculated via auto-differentiation. The cost function is once again Equation \ref{eq:cost_function}.

For the gradient step, unlike that of $\theta$, the values of amplitude display a much broader dynamic range (several orders of magnitudes). So for optimal convergence, it is necessary to have a step size that varies from pixel to pixel and over iterations as a more accurate guess of the amplitude is obtained. Fortunately, in this case, the second derivative is explicit:
\begin{equation}
H_{A^c}=\frac{\partial^2 \mathcal{L}}{\partial A_{(i,j)}^{c \;2}}=\sum_{(i,j,k)}^{n_P,n_E}\frac{\mathcal{M}^{c\; 2}(\theta_{(i,j,k)}^c)X_{(i,j,k)}}{\sum_{x=0}^{n_C}\Big(A_{(i,j)}^x\mathcal{M}^x(\theta_{(i,j,k)}^x)\Big)},
\end{equation}
so it does not cost much to calculate the diagonal elements of the Hessian $H_{A^c}$. Thus, at each iteration, the gradient step size is calculated as
\begin{eqnarray}
    \alpha_{A_c}=1/H_{A^c}.
\end{eqnarray}
Finally, we imposed that the amplitude should be non-negative, so if after the gradient descent step $A<0$, we put $A\leftarrow 0$.
\subsubsection{Stopping criterion and output}

Values of the cost functions are recorded in the list $L$ of length $p$, the number of iterations. The iterations stop when the following criterion is reached, which evaluates an average of how much the cost function has fluctuated in the past  fifty iterations:
\begin{equation}
    \frac{
    \overline{\Big\{L_{p-150}-L_{p-50},...,L_{p-100}-L_{p}\Big\}}}{\overline{\Big\{L_{p-150},...,L_{p-100}\Big\}}}<\epsilon
    \label{eq:stop_criterion},
\end{equation}
where $\epsilon$ is a chosen small value, such as $10^{-6}$.
The output are fitted hyperspectral cubes $\{\hat{X}^0,...\hat{X}^C\}$, one for each component and their sum, and the total fit $\hat{X}$.
If the stopping criterion is never reached, a maximum number of iterations can be set (in our tests, a good choice for the maximum was of the order of $10^4$ iterations).
\begin{algorithm}
\caption{SUSHI: Semi-blind Unmixing with Sparsity for Hyperspectral Images.}\label{alg:sushi}
\begin{algorithmic}
\State \textbf{input} data X, trained IAE models $\{\mathcal{M}^0,...,\mathcal{M}^{n_C}\},$ number of wavelet scales $J$, sparsity threshold factor $k$, cost function $\mathcal{L}$.
\State \textbf{initialization} $\{\theta^0_0,...,\theta^{n_C}_0\} \leftarrow \{\mathbb{1}/N_A^0,...,\mathbb{1}/N_A^{n_C}\} $
\State $\{A^0_0,...,A^{n_C}_0\} \leftarrow \sum_{e}^{n_E} X(.,e)/n_C $
\State $\alpha_{\theta} \leftarrow  0.1 / \max(A^0_0)$
\State $t \leftarrow 0$
\While{stopping criterion \ref{eq:stop_criterion} is not met}
\For{component $c$ in $\{0,...,n_C\}$}
\State \textbf{Gradient descent step on $\theta^c$}
\State $\theta^c_{t+1/2} \leftarrow \theta^c_{t} - \alpha_{\theta} \nabla_{\theta^c} \mathcal{L}(\theta^c|X,A^c,\theta^{C\neq c})$
\State \textbf{Sparsity proximal step on $\theta^c$}
\State $\theta^c_{t+1} \leftarrow $ \textbf{prox}$_{l_1,J,k}(\theta^c_{t+1/2})$ (using Algorithm \ref{alg:proxL1})
\State \textbf{Gradient descent step on $A^c$}
\State $H \leftarrow \nabla^2_{A^c}(\mathcal{L}(A^c|X,\theta^c_{t+1}))$ 
\State $A^c_{t+1/2} \leftarrow A^c_{t} - 1/H \nabla_{A^c} \mathcal{L}(A^c|X,\theta^c_{t+1})$
\State \textbf{Non-negativity proximal step}
\State $A^c_{t+1} \leftarrow A^c_{t+1/2} \text{ where } A^c_{t+1/2}>0, \text{and } 0$ elsewhere.
\EndFor
\State $t \leftarrow t+1$
\EndWhile
\State$\hat{X}^c \leftarrow A^c_t\mathcal{M}^c(\theta^c_t)$
\State$\hat{X} \leftarrow \sum_{c=0}^{n_C} \hat{X}^c$
\State \Return $\hat{X}, \{\hat{X}^0,...\hat{X}^C\}$
\end{algorithmic}
\end{algorithm}

\begin{algorithm}
\caption{Proximal Operator of the L1 norm.}\label{alg:proxL1}
\begin{algorithmic}
\State \textbf{input} image $theta$, number of wavelet scales $J$, sparsity threshold factor $k$,
\State \textbf{functions}
\State - WT: starlet transform.
\State - prox: proximal operator \textbf{from equation \ref{eq:proxl1}}.
\State - mad: median absolute deviation.
\State $\{w_c, w_0,...,w_J\} \leftarrow \text{WT}(\theta^c_{t+1/2})$
\State $\{w_c, wg_0,...,wg_J\} \leftarrow \text{WT}(\theta^c_{t+1/2})$
\For{j in \{0,...,J\}}
    \State $\rho_j \leftarrow \textbf{mad}(wg_j) $
    \State $w_j \leftarrow \textbf{prox}(w_j,k\rho_1)$
\EndFor
\State $\hat{\theta}\leftarrow w_c + \sum_{j=0}^{J} w_j$
\Return $\hat{\theta}$
\end{algorithmic}
\end{algorithm}

\section{Results on simulated data}
\label{sec:results_sim}
In this section, we present the toy model used to test our algorithm. We then compare the results obtained by SUSHI to the classic method (a pixel-by-pixel fit with no regularization).

\subsection{Toy model}
\label{sect:toymodel}
\paragraph{}
The toy model used to test SUSHI was inspired by case studies of X-ray imagery of supernova remnants, such as those taken by the Chandra telescope. The purpose of our use case here is to map the properties of the hot ejecta and the synchrotron component using only data above 3 keV (focusing on the continuum and the Fe-K line complex).
Thus, the toy model has two physical components. The first represents the thermal emission of collisionally ionized hot gas, for which the spectral model is APEC.\footnote{\url{http://www.atomdb.org/physics.php}} The second represents synchrotron emission from electrons accelerated at the supernova forward shock, for which the spectral model is a power law. These were convolved by the Chandra instrumental response (ARF and RMF files).

The amplitude maps of the two components can be found in Section \ref{sec:results_sim}, where they are compared to the results obtained by SUSHI and the classic method. The synchrotron amplitude map came from real Cassiopeia A data analyzed by the stationary unmixing technique described in \cite{Adrien_GMCA}. The amplitude map of the thermal component, on the other hand, was taken from a numerical simulation by \cite{Orlando_2016} that was meant to be similar to the ejecta in Cassiopeia A.

Apart from amplitude, the spectra for the toy model's thermal component were temperature (kT) and velocity redshift (z). The maps for these parameters, which we used to generate the toy model's spectra, were also taken from the \cite{Orlando_2016} simulation and can also be found in Section \ref{sec:results_sim}. For the sake of simplicity, only one temperature and one velocity was simulated for each line of sight,\footnote{For each line of sight, the average parameter was obtained by weighting by X-ray emissivity .} and only a half-sphere of the remnant was taken (the one moving toward the observer). As for the synchrotron component, the only physical parameter apart from amplitude is the photon index, and we opted to keep it constant at a value of 2.5 for simplicity, which is similar to the values found in Cassiopeia A in X-rays.

\begin{figure}[H]
    \centering
    \includegraphics[width=\linewidth]{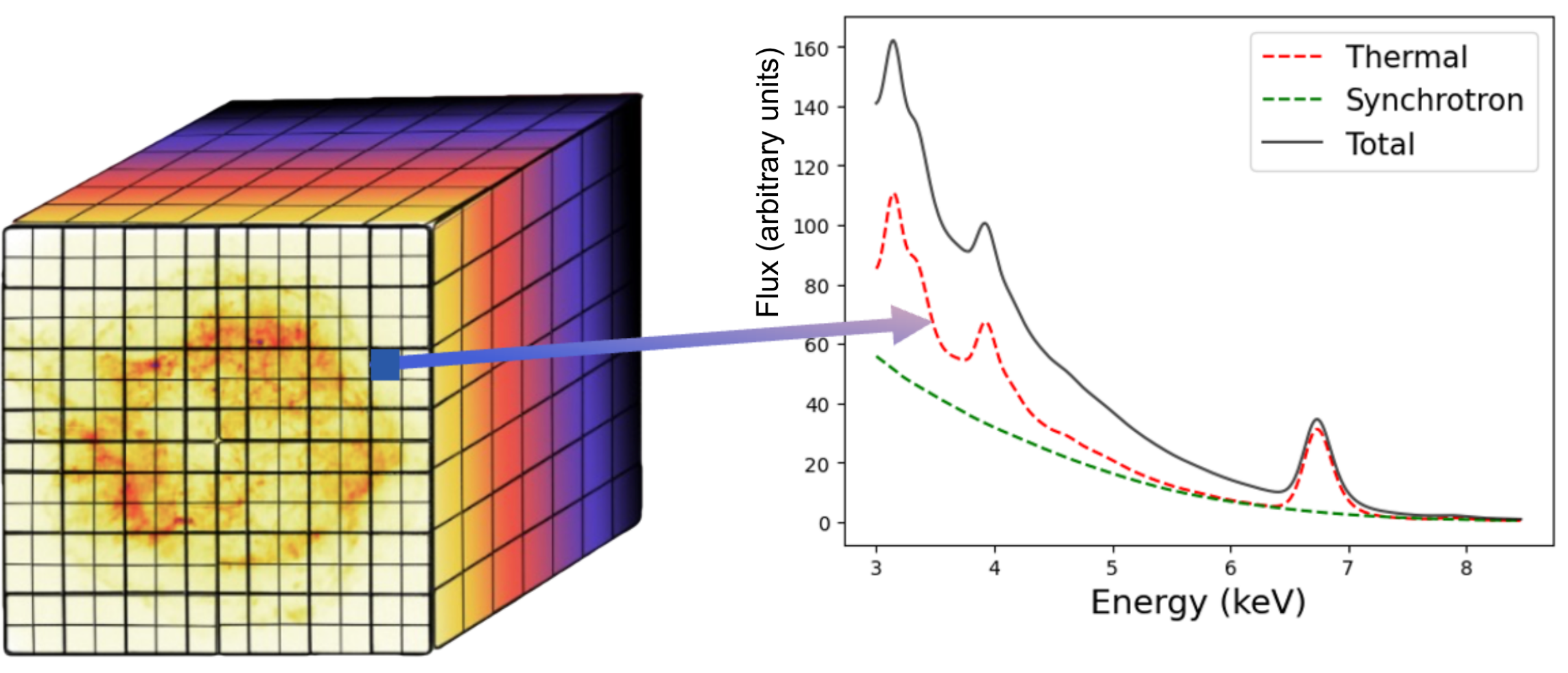}
    \caption{Diagram showing the hyperspectral structure of the toy model. On the left is a schematic depiction of a hyperspectral cube. On the right is an example of a spectra for one pixel, with the two components in color and their sum in black.}
    \label{fig:toymodel_diagram}
\end{figure}

The resulting toy model has 375 energy channels between 3 keV and 8.48 keV and 94$\times$94 pixels, resulting in a cube of size 375$\times$94$\times$94. A diagram is available in Figure \ref{fig:toymodel_diagram}. Upon this ground truth, we applied Poisson noise to generate the data set. Different noise levels were generated to test the performance of SUSHI at various regimes. This work presents the results for noise level regimes similar to the real data of Cassiopeia A observed in 2004 for 980 ks in the same energy band.

\subsection{Results}
\paragraph{}
This section presents the results obtained by the SUSHI method when performing non-stationary unmixing on the toy model described in the previous section. Two trained IAEs were used as our surrogate spectral models, one for the synchrotron component, with $n_A=2$, trained on power laws convolved by the Chandra instrument response, and the other for the thermal component, with $n_A=4$, trained on simulated thermal emission spectra
of collisionally ionized hot gas also convolved by the instrument response. More details about the architecture of the IAEs can be found in the appendix, Section \ref{sec:Appendix_IAE}.

Our point of comparison is a classic multivariate pixel-by-pixel fit obtained by the Xspec\footnote{\url{https://heasarc.gsfc.nasa.gov/xanadu/xspec/}} fitting package, 
which contains the same physical models as those used to train our surrogate models. This method is dubbed the "classic method." \\
At this point in time, SUSHI's final output is a fitted unmixed spectra, but it does not provide physical parameters (though this is an area of further work we intend to explore, it is outside the scope of this paper). Thus, to obtain parameter maps, the de-noised unmixed spectra obtained by SUSHI were fitted with Xspec at the end of the process in order to retrieve physical parameters.

\begin{table*}
    \centering
    \begin{tabular}{|c|c|c|c|c|}
    \hline
         \textbf{Parameter} & \textbf{SUSHI residual mean} & \textbf{Classic residual mean} & \textbf{SUSHI residual std} & \textbf{Classic residual std}\\
         \hline
         Thermal Amplitude & -0.094 & -0.086 & 1.01 & 1.06\\
         \hline
         Synchrotron Amplitude & -0.29 & -0.58 & 5.17 & 8.56\\
         \hline
         Temperature (kT) & -0.05 & -0.23  & 0.18  & 2.77  \\
         \hline
         Velocity Redshift (z) & -0.14 & -0.06 & 1.25 & 9.60 \\
         \hline
         Photon index ($\Gamma$) & -0.02 & -0.16 & 0.06 & 0.65 \\
         \hline
    \end{tabular}
    \caption{Table summarizing the means and standard deviations (std) of the fitted parameter residuals. For a parameter with ground truth $\tilde{x}$ and fit $\hat{x}$, the residual of $x$ is defined as $(\tilde{x} - \hat{x})/\tilde{x}$. The lower the absolute value, the better (perfect fit having a zero mean and standard deviation).}
    \label{tab:res_mean_std}
\end{table*}
\begin{figure*}
    \centering
    \includegraphics[width=0.9\textwidth]{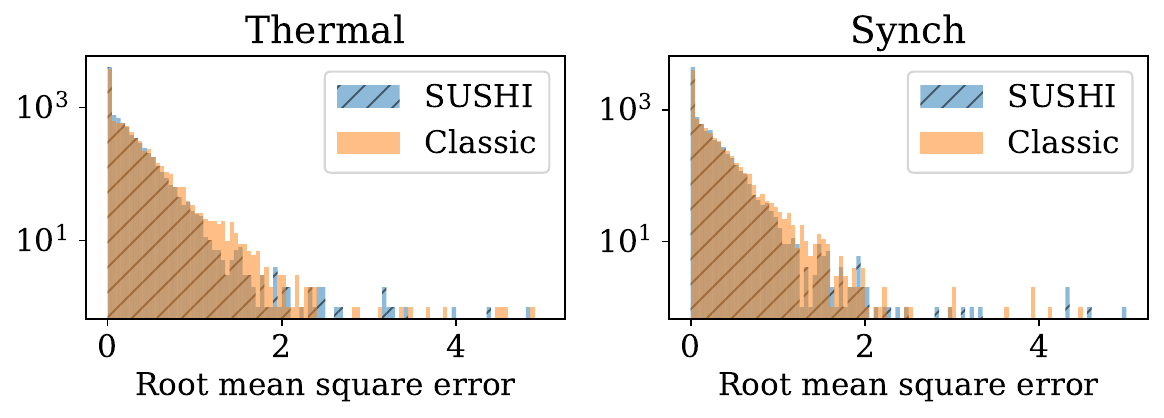}
    \caption{Root mean squared error histogram for each spectrum. For each component $c$ and pixel $(i,j)$, RMSE$^c(i,j)=\sqrt{\sum_k^{n_E}(\tilde{X}^c(i,j,k)-\hat{X}^c(i,j,k))^2/l}.$}
    \label{fig:RMSE_hist}
\end{figure*}

\begin{figure*}
    \centering\includegraphics[width=0.9\textwidth]{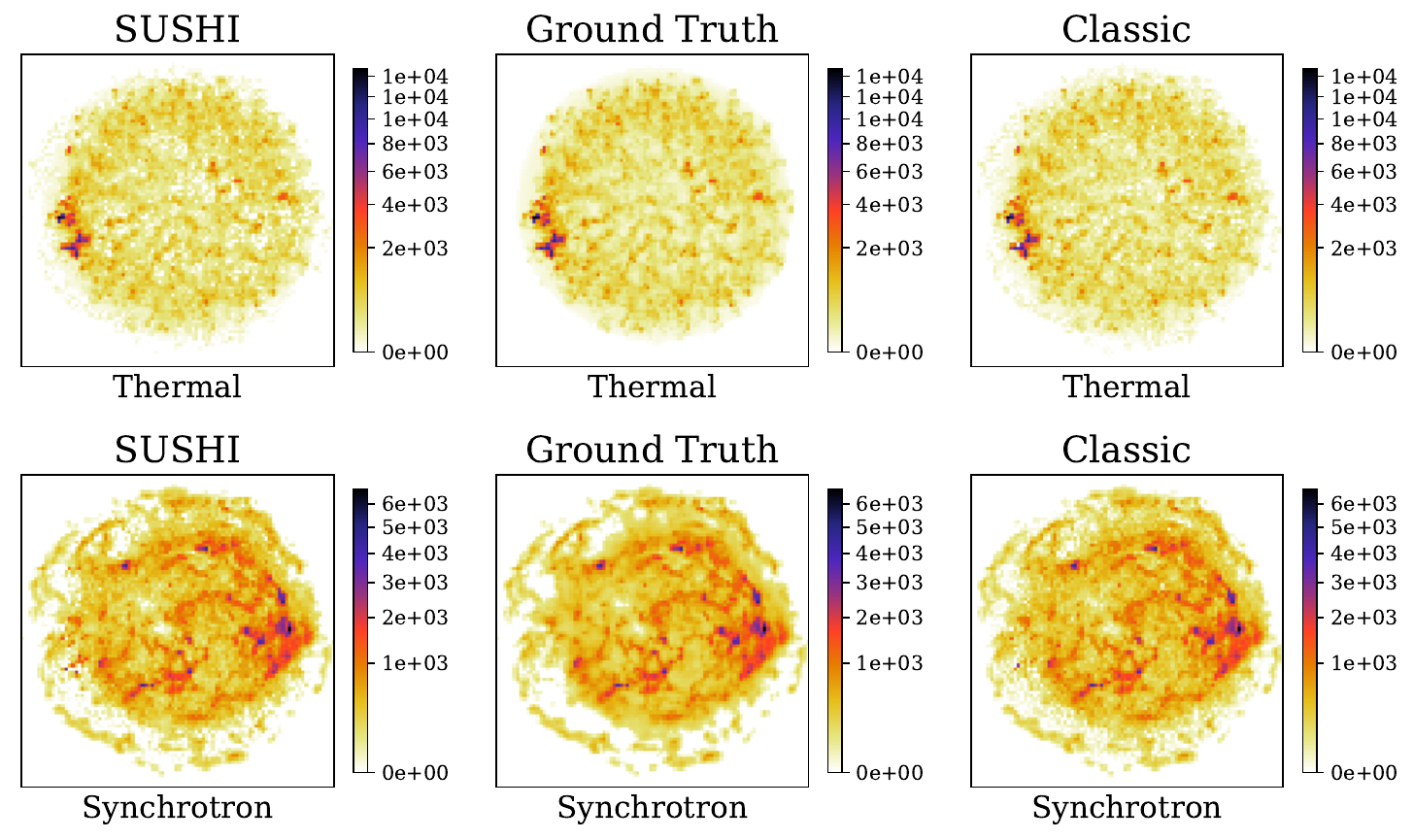}
    \caption{Obtained amplitude maps for each component and each technique compared to the ground truth. The color represents 
    the level of brightness (the total number of counts per pixel over the energy band considered).}
    \label{fig:Amp_map}
\end{figure*}

Figure \ref{fig:RMSE_hist} shows the histograms of the root mean squared error of the hyperspectral reconstructed cube for the two components. We find that SUSHI has a somewhat smaller error for both components. For SUSHI, the average root mean squared error for each spectra was found to be 0.19 for the thermal and 0.16 for the synchrotron, while for the classic method it was 0.22 for the thermal and 0.19 for the synchrotron.

\begin{figure*}[h!]
    \centering\includegraphics[width=0.9\textwidth]{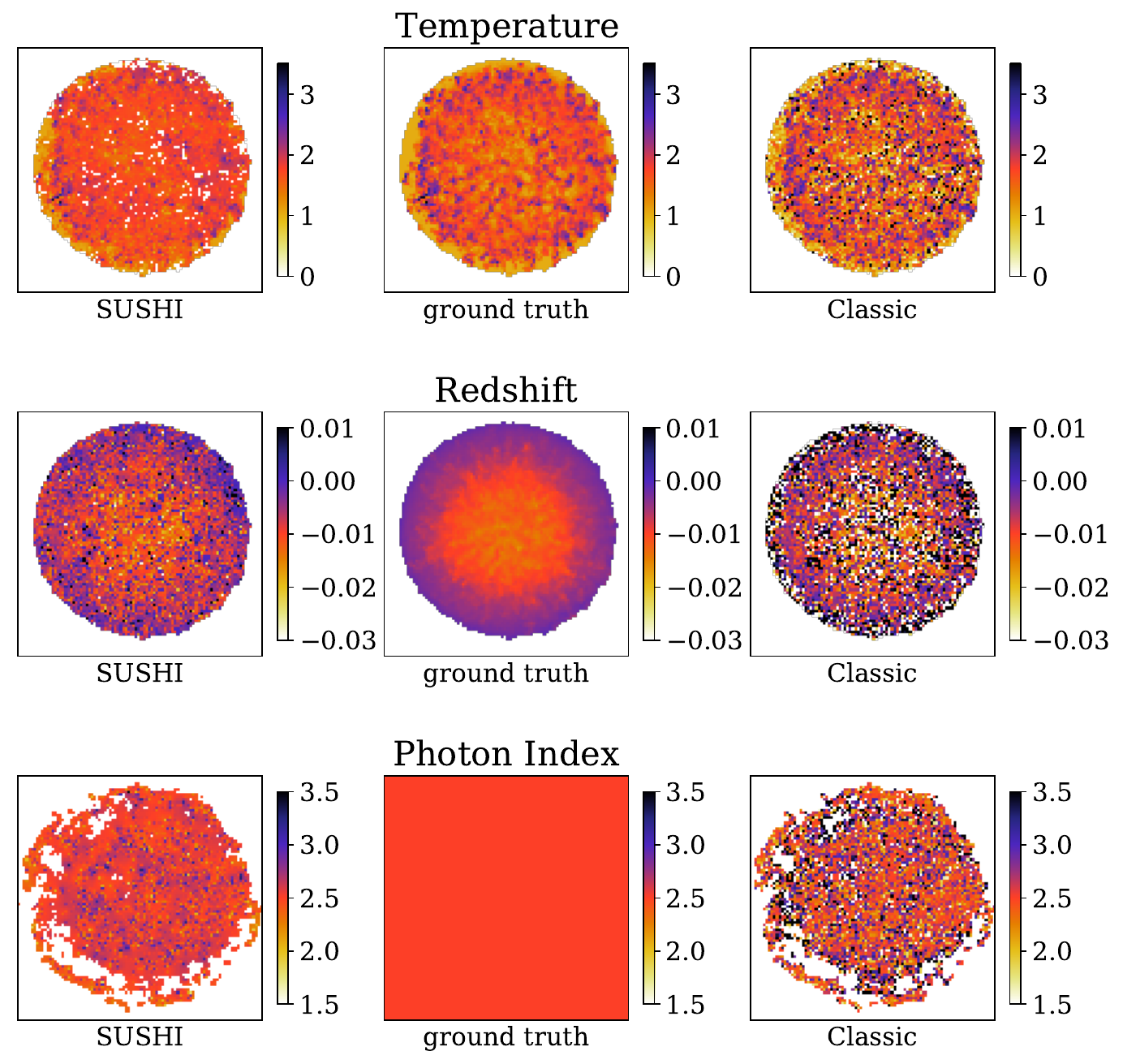}
    \caption{Obtained physical parameter maps using SUSHI and the classic method compared to the ground truth of the toy model. }
    \label{fig:tm_param_maps}
\end{figure*}

\begin{figure*}[h!]
    \centering
    \includegraphics[width=0.9\textwidth]{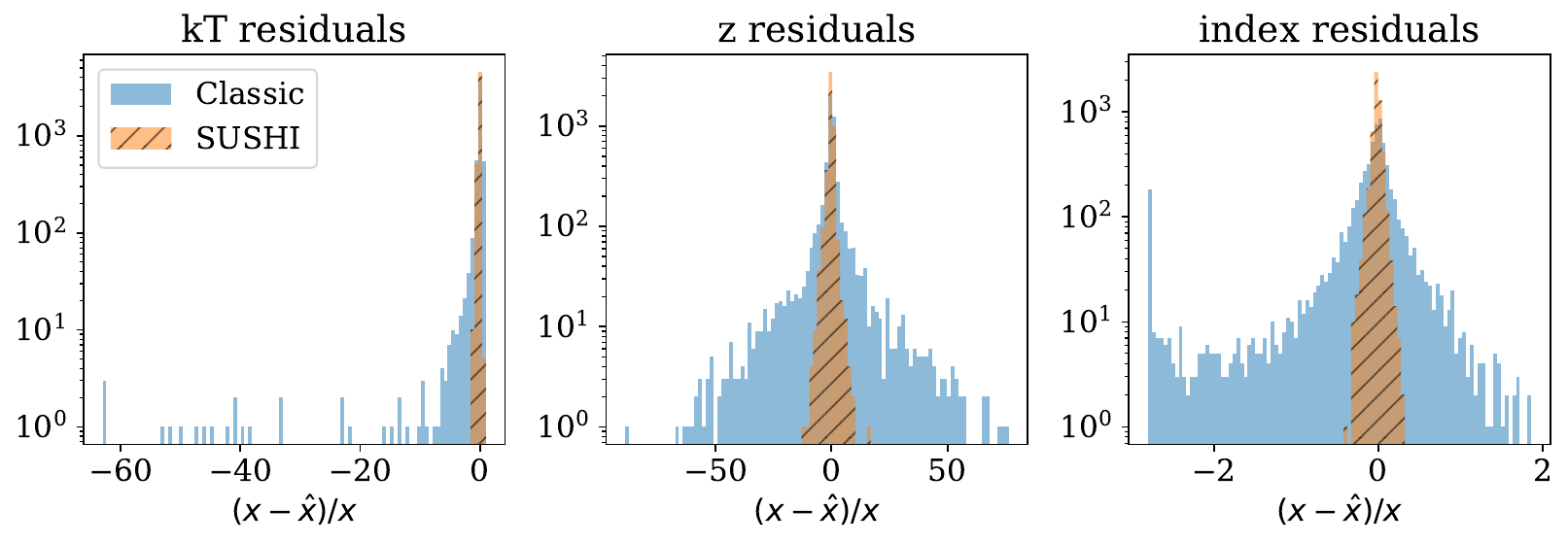}
    \caption{Histograms of the toy model parameter residuals. For a parameter with ground truth $\tilde{x}$ and fitted value $\hat{x}$, the residual of $x$ is defined as $(\tilde{x} - \hat{x})/\tilde{x}$. The lower the absolute value, the better (perfect fit having a zero mean and standard deviation). }
    \label{fig:paramreshist}
\end{figure*}

\begin{figure*}
\begin{subfigure}{0.6\textwidth}
  \hspace*{3.5cm}\includegraphics[height=0.23\textheight]{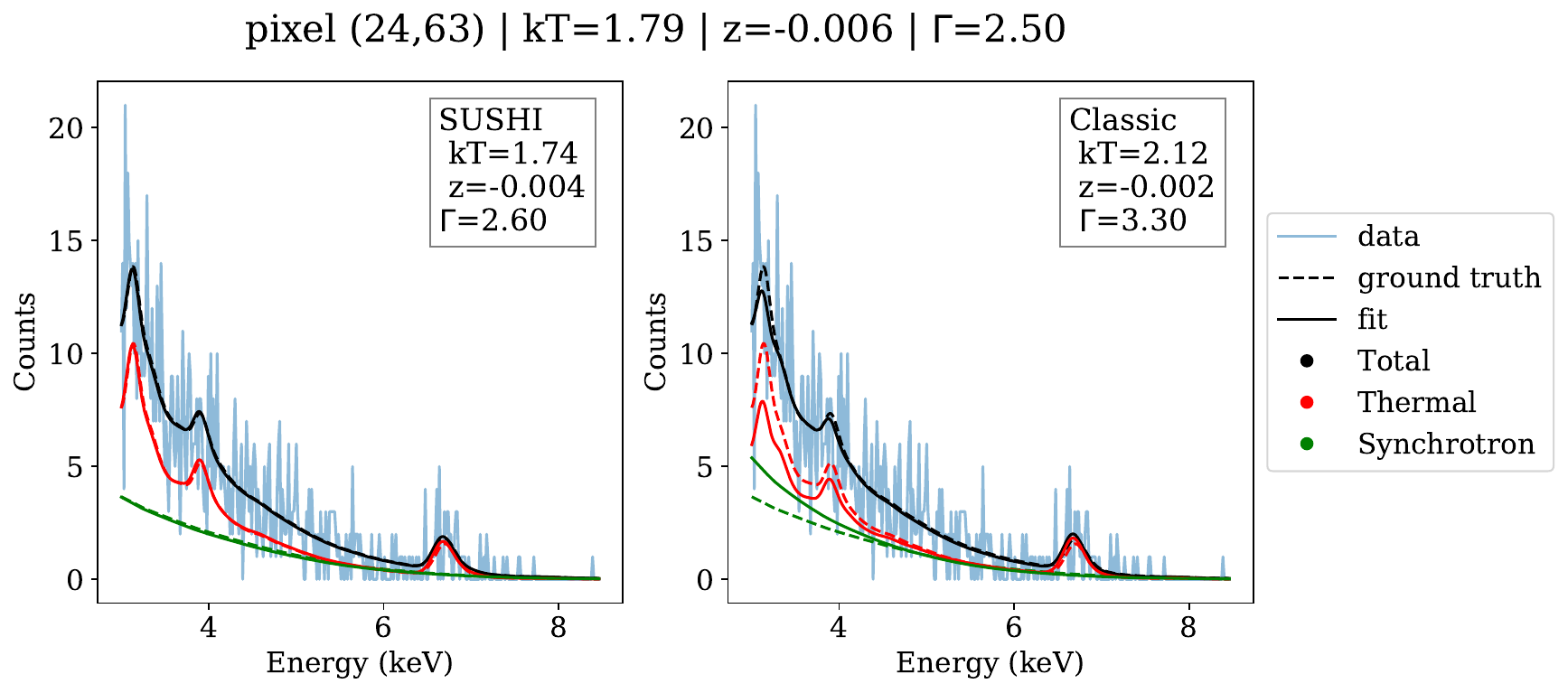}
\end{subfigure}%
\\
\begin{subfigure}{0.6\textwidth}
  \hspace*{3.5cm}\includegraphics[height=0.23\textheight]{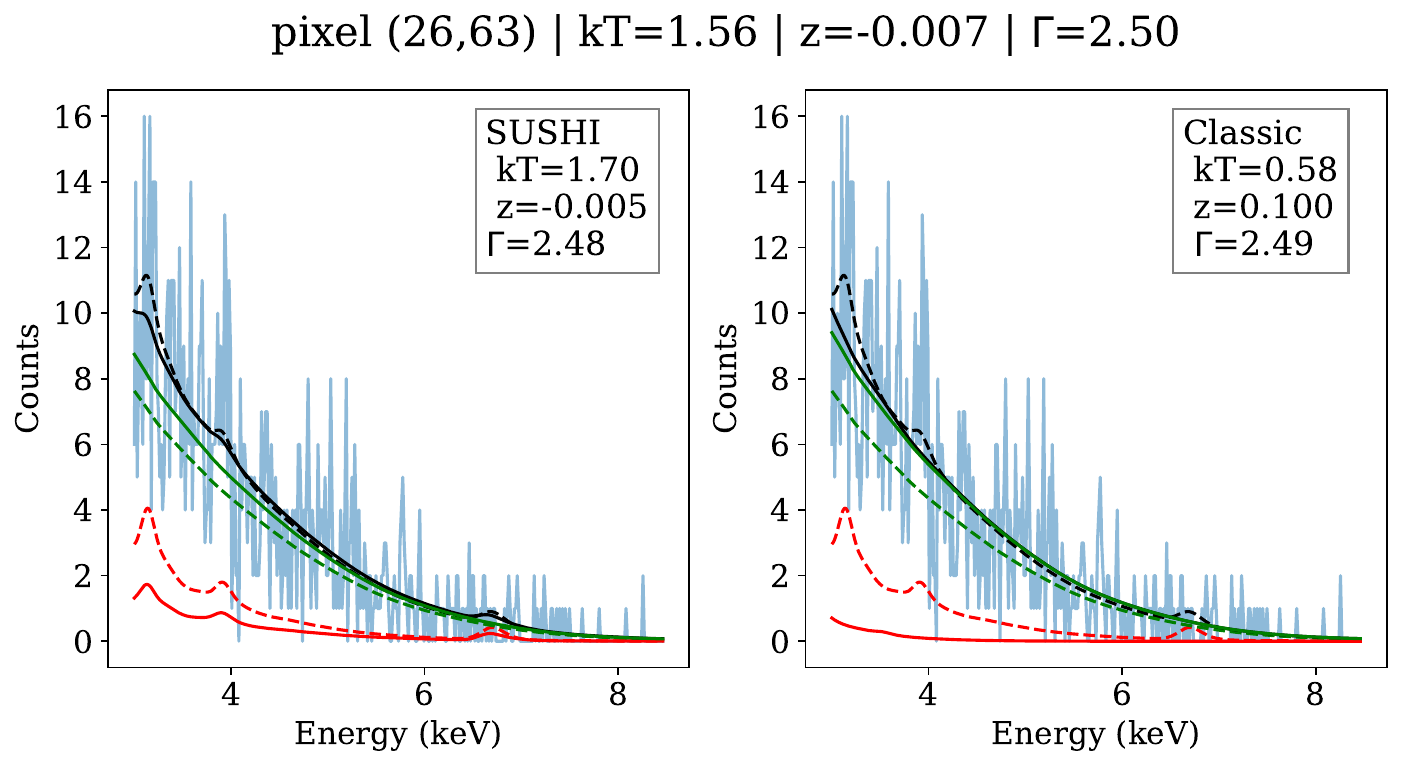}
\end{subfigure}
\\
\begin{subfigure}{0.6\textwidth}
  \hspace*{3.5cm}\includegraphics[height=0.23\textheight]{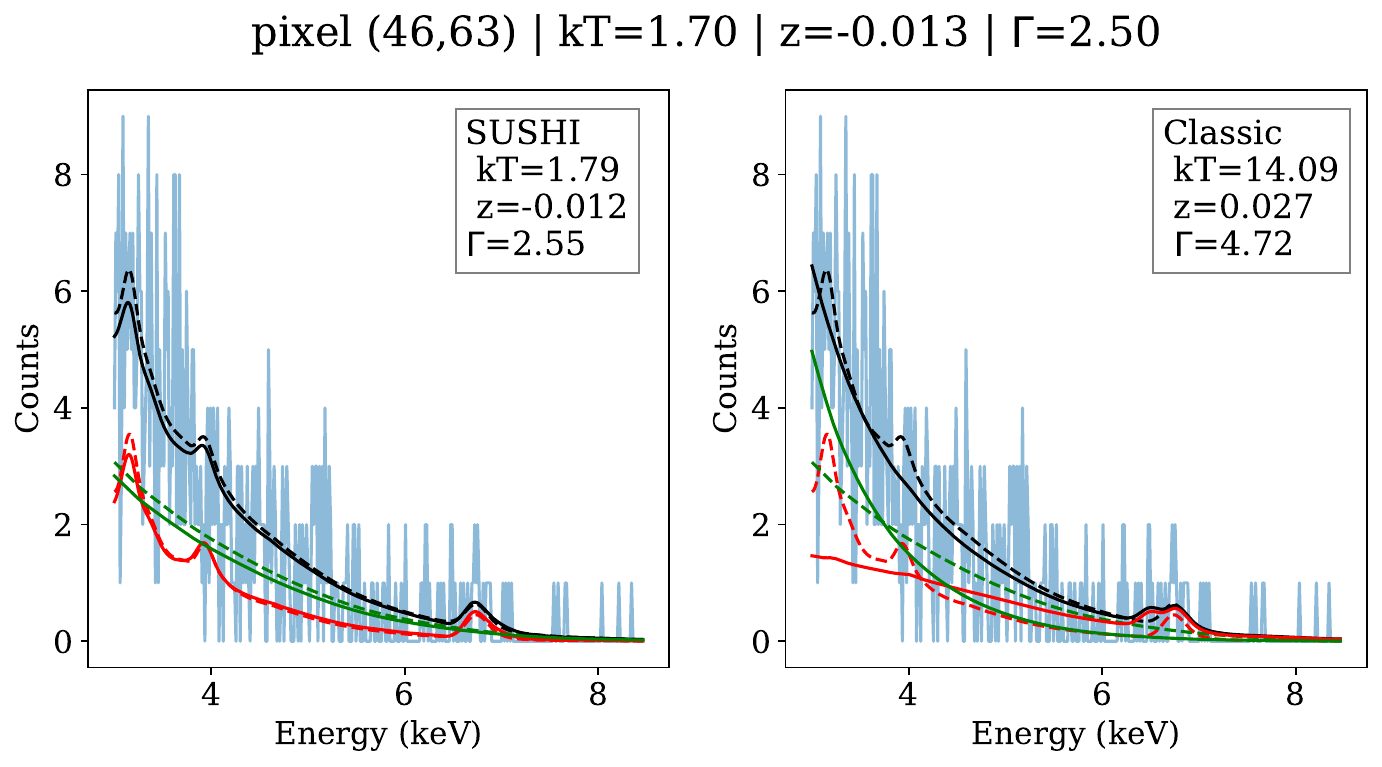}
\end{subfigure}%
\\
\begin{subfigure}{0.6\textwidth}
  \hspace*{3.5cm}\includegraphics[height=0.23\textheight]{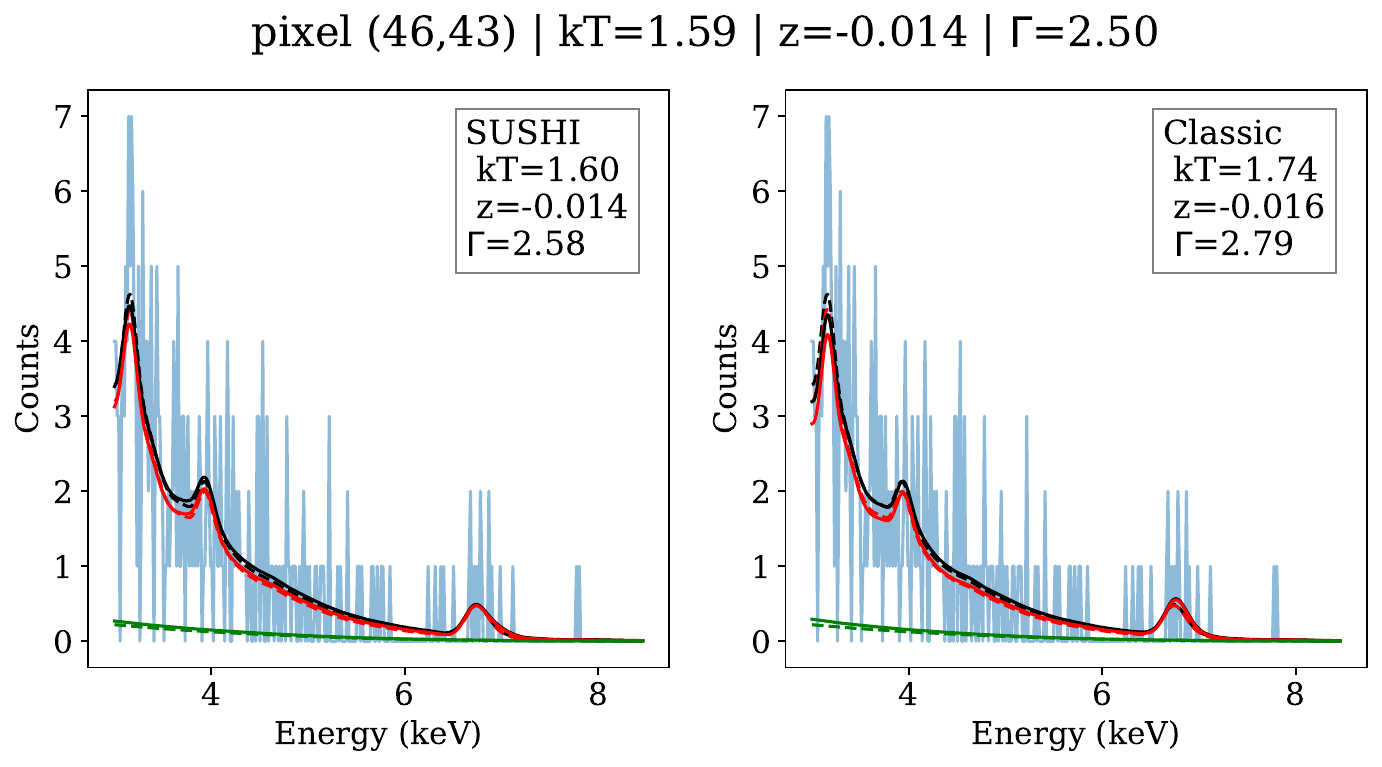}
\end{subfigure}
    \caption{Examples of the fit performed by SUSHI (left) versus the classic method (right) for four pixels from the toy model. In the title are written the pixel coordinates and the ground truth parameters (temperature kT in keV, redshift z, photon index $\Gamma$) to be compared with the best-fit parameters in the legend of each graph.}
    \label{fig:pixelexamples}
\end{figure*}

Figure \ref{fig:Amp_map} shows a comparison of the obtained amplitude maps for the two components. We observed that both methods succeed reasonably well in finding the amplitude of both components. Looking at Table \ref{tab:res_mean_std}, which gathers the mean and standard deviation of its residuals with the ground truth for both methods and every reconstructed parameter (a perfect fit would have a mean and standard deviation close to zero), we found that SUSHI is as good as the classic method for the thermal amplitude but better than the classic method for the synchrotron amplitude.

However, the classic method falls short when it comes to parameter mapping.
Figure \ref{fig:paramreshist} shows a histogram of the residuals on the three physical parameters. We notice that SUSHI obtains a much better reconstruction of the physical parameters. Figure \ref{fig:tm_param_maps} provides a comparison of the obtained best-fit physical parameter maps.
Based on Table \ref{tab:res_mean_std}, we observed that SUSHI outperforms the classic method for every fitting parameter except the thermal amplitude, where the two methods are equally accurate. For temperature and the photon index in particular, the standard deviation is reduced by an order of magnitude.

Some examples of reconstructed pixels can be seen in Figure \ref{fig:pixelexamples}. On pixel (24,63), one can see a case where the classic method underestimated the thermal component and overestimated the synchrotron at lower energies, whereas SUSHI was able to avoid this misstep. In pixel (26,63), the task of reconstructing the thermal component is especially difficult, as the thermal peaks are all drowned out by the synchrotron component, but SUSHI retrieved better results than the classic method. Pixel (46,63) presents a similarly difficult case, and SUSHI was able to achieve a very good fit. Finally, pixel (46,43) shows an example where the reconstruction is easy, and both methods achieved a good fit.
Thus, we find that SUSHI manages to reconstruct the hyperspectral cube much better than the classic method, even on pixels with low signal-to-noise ratios or those where one of the components is drowned out.

\section{Application to real X-ray data}
\label{sec:results_data}
After benchmarking the method on a simulated data set, we explored the performance of the method on real data: a hyperspectral image of the Cassiopeia A SNR, as taken by the Chandra X-ray telescope.
In this section, SUSHI is used to analyze a real data set: a hyperspectral image of the Cas A SNR and of the Crab pulsar wind nebula, each taken by the Chandra X-ray telescope.
\subsection{The Cassiopeia A supernova remnant}
\label{sec:results_CasA}
\begin{figure}
\centering
  \includegraphics[width=0.8\linewidth]{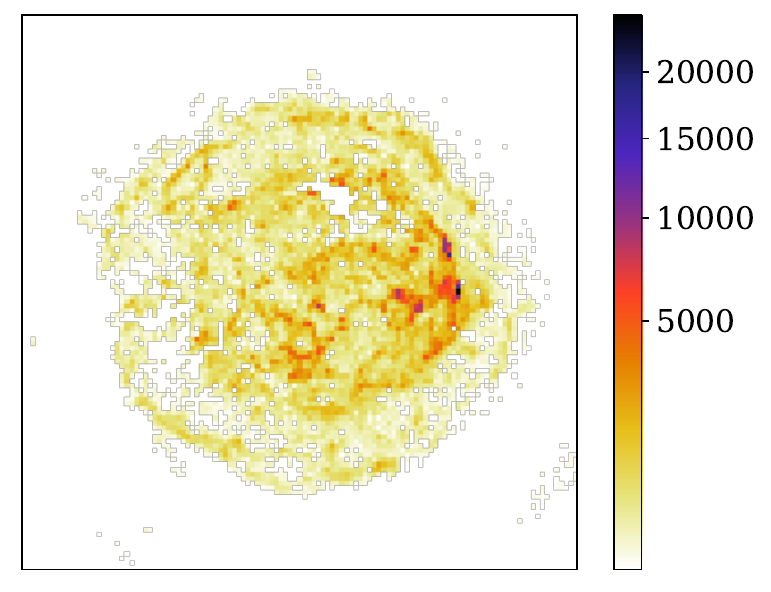}
  \caption{Retrieved amplitude map for the synchrotron component of the Cassiopeia A data set. The color represents the total number of counts per pixel over the 4.21-7.48 keV energy band.}
  \label{fig:CasA_Ampl_Synch}
\end{figure}%
\begin{figure}
  \centering
  \includegraphics[width=0.8\linewidth]{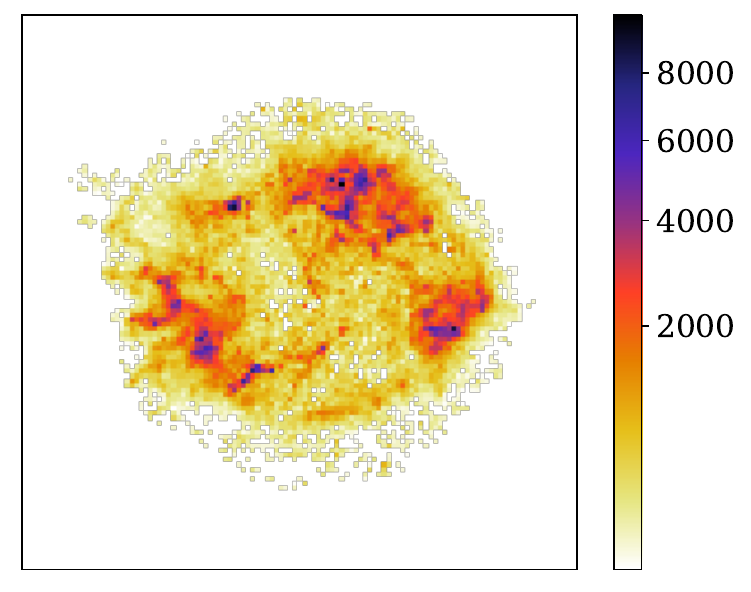}
\caption{ Retrieved amplitude map for the thermal component of the Cassiopeia A data set.
The color represents the total number of counts per pixel over the energy band considered.}
\label{fig:CasA_Ampl_Therm}
\end{figure}
\begin{figure}
\centering
  \centering
  \includegraphics[width=0.8\linewidth]{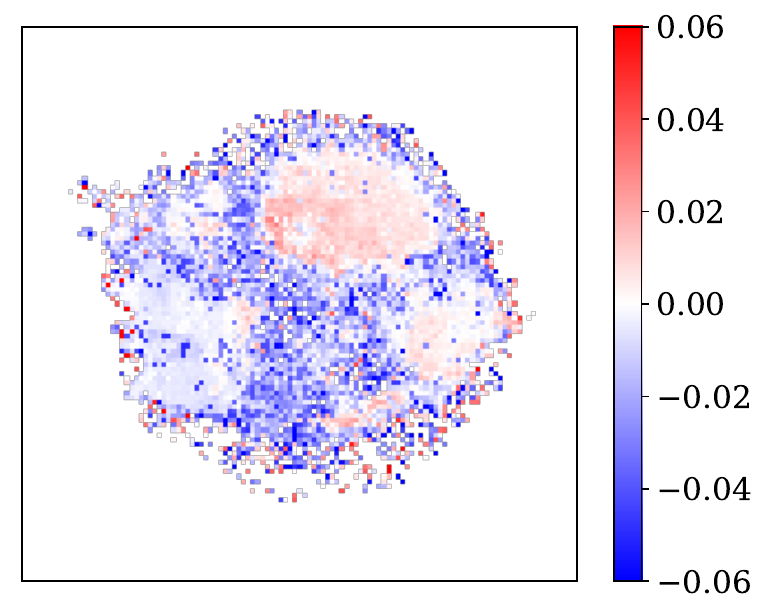}
  \caption{Velocity redshift map for the Cassiopeia A data set.}
  \label{fig:CasA_z}
\end{figure}%
\begin{figure}
  \centering
  \includegraphics[width=0.8\linewidth]{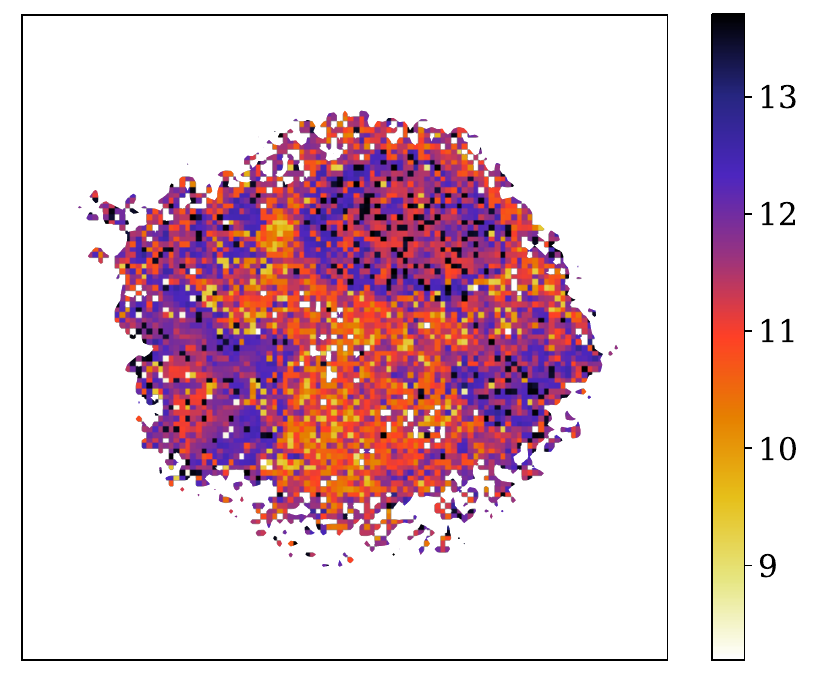}
  \caption{Ionization timescale map $log_{10}$(cm$^{-3}$ s) for the Cassiopeia A data set.}
  \label{fig:CasA_tau}
\end{figure}
\begin{figure*}
\centering
\begin{subfigure}{.4\linewidth}
  \centering
  \includegraphics[width=\linewidth]{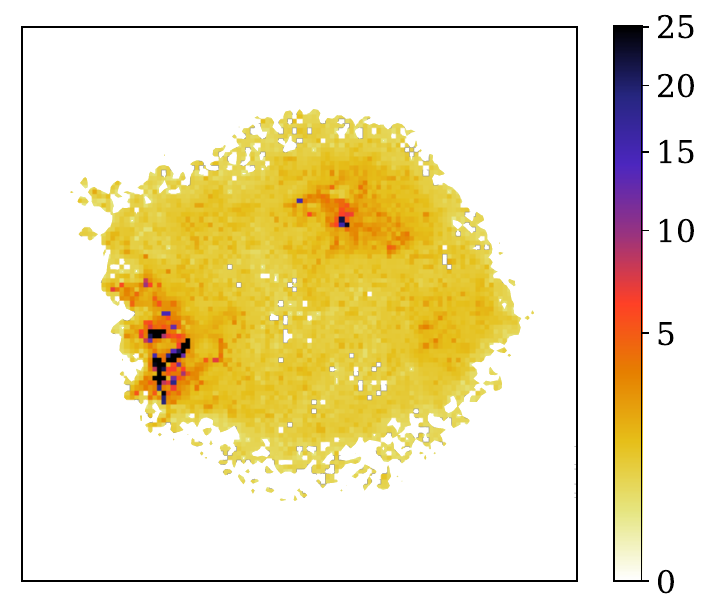}
  \caption{}
\end{subfigure}%
\begin{subfigure}{.4\linewidth}
  \centering
  \includegraphics[width=\linewidth]{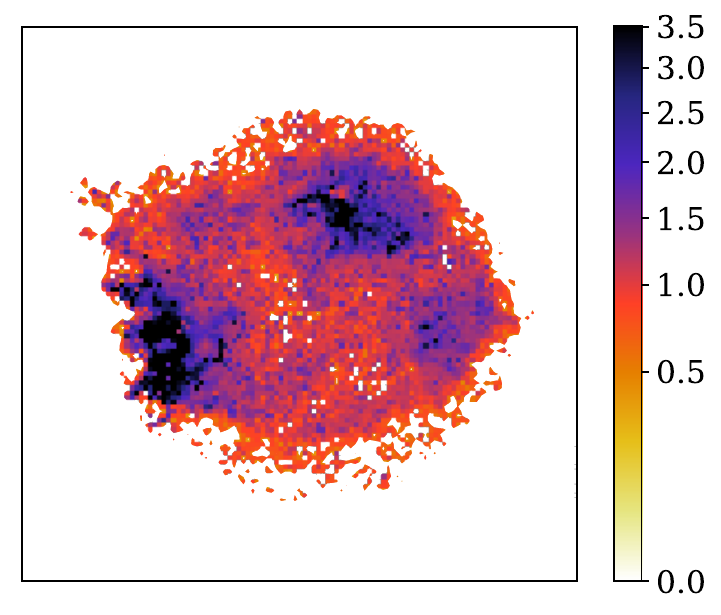}
  \caption{}
\end{subfigure}
\caption{Temperature map (keV) for the Cas A data set. (a) Temperature map with no boundaries on the color map. (b) Temperature map where the maximum of the color map has been restrained to 3.5 keV.}
\label{fig:CasA_kT}
\end{figure*}

\begin{figure*}
\centering
\begin{subfigure}{0.4\linewidth}
  \centering
  \includegraphics[width=\linewidth]{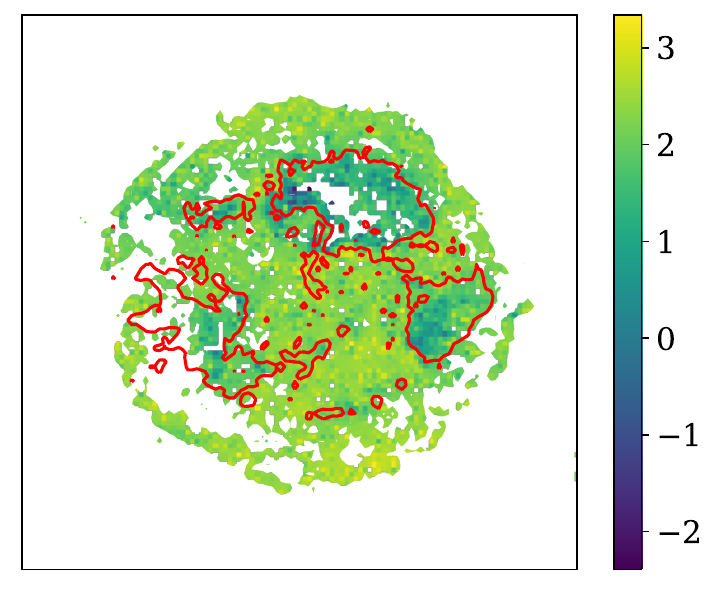}
  \caption{}
\end{subfigure}%
\begin{subfigure}{0.4\linewidth}
  \centering
  \includegraphics[width=\linewidth]{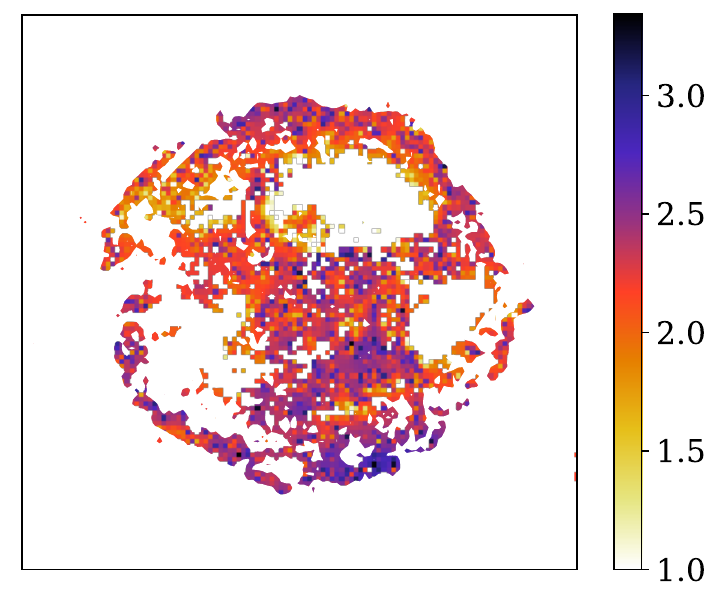}
  \caption{}
  \label{fig:CasA_pho}
\end{subfigure}
\caption{
Reconstructed synchrotron photon index map. (a) Inside the red contours, the amplitude of the thermal component was over $10^3$ per pixel. In this area, there was on average over twice as much of the thermal component (the ratio of the mean thermal amplitude over the mean synchrotron amplitude was 2.2).
(b) Same photon index map as (a) but a minimum value of the index was fixed at one.
We observed that though our algorithm has found some negative values for the photon index in some pixels, they are only in areas where the thermal component drowns out the synchrotron component completely and can thus be disregarded.}
\end{figure*}

Cassiopeia A SNR results from a core collapse supernova and is one of the youngest remnants in our Galaxy, with an estimated age of $\sim$ 350 yrs. Being the brightest extended source in X-rays, it has been extensively studied in the literature \citep[see ][ and references therein for example]{Picquenot21}.
The data set we chose is the same set that \cite{Adrien_GMCA} analyzed using the stationary unmixing technique GMCA.
It comprises observations of the Cassiopeia A SNR as taken by the Chandra ACIS-S instrument in 2004, with a total exposure time of 980 ks, that were merged into a single hyperspectral cube. The spatial binning is 2'', the spectral binning is 14.6 eV, and an energy range between 4.21 keV and 7.48 keV was chosen.

For the sake of simplicity, no background emission was included in our toy model described in Section \ref{sect:toymodel}. However, to apply our method to real data, we needed to account for the instrumental and astrophysical background.
To account for this background emission in our method, we added a linear term $B\in {\mathbb{R}^{n_E}}$ to the model in our cost function (equation \ref{eq:fullproblem_regl1_poisson}) that was constant for every pixel.
This background spectrum was calculated using the blank-sky\footnote{\url{https://cxc.cfa.harvard.edu/ciao/threads/acisbackground/}} observations provided by the Chandra ACIS calibration team.
This background hyperspectral image was not modified during our fitting process. 

Since reality is more complex than our toy model, another IAE model, with $n_A=6$,  was trained for the thermal component to handle the fact than in Cassiopeia A, the plasma is not in ionization equilibrium.
The training set were instances of emission spectra from a non-equilibrium ionization collisional plasma model (NEI\footnote{\url{https://heasarc.gsfc.nasa.gov/xanadu/xspec/manual/node195.html}} model in Xspec).
The variable parameters for the thermal model were temperature kT, the velocity redshift z, and the ionization timescale $\tau$.
The metal abundance of elements was kept constant.
The final model was composed of the aforementioned thermal component and the synchrotron power-law component.

\paragraph{}
The retrieved amplitude maps are shown in Figures \ref{fig:CasA_Ampl_Synch} and \ref{fig:CasA_Ampl_Therm} for the synchrotron and thermal components, respectively. The synchrotron component displays a filamentary structure, while the thermal component shows the supernova remnant ejecta, dominated here by the Fe-K emission line at 6.5 keV. When comparing with the results obtained by \cite{Adrien_GMCA}, who performed a stationary unmixing on the same data set, we observed that their results are consistent with ours: The synchrotron amplitude is similar, and the thermal amplitude looks like the sum of the two thermal components found by \cite{Adrien_GMCA}, which had been interpreted as redshifted and blueshifted ejecta.

In Figure \ref{fig:CasA_z}, we show the velocity redshift map. When looking at particular areas in the upper right and lower left of the map, which correspond in Figure \ref{fig:CasA_Ampl_Therm} to high-amplitude areas, we observed that the upper clump is redshifted, while the lower clump is somewhat blueshifted. This is in accordance with the asymmetry found by \cite{Adrien_GMCA}, but now much more detail is present regarding the redshift variability. The rest of the map corresponds to areas where the amplitude of the thermal component is much lower, and we observed a tendency toward a blueshift.

Figure \ref{fig:CasA_tau} and Figure \ref{fig:CasA_kT} show the ionization timescale and the temperature maps of the thermal component,
respectively. We noticed that high temperatures and high ionization are often in similar locations, in particular in the aforementioned "clumps."
In Figure \ref{fig:CasA_pho} is the map of the obtained photon index of the synchrotron component. Notably, this map helped us see a limitation of our model, namely that we kept the abundance of iron constant, and the best fit uses the synchrotron component to compensate for this fact. Indeed, there are areas where the synchrotron photon index is negative, which is not a physically realistic result. Those areas correspond to those with a high thermal amplitude and high temperature.

\begin{figure}
\centering
\begin{subfigure}{\linewidth}
  \centering
    \includegraphics[width=\textwidth]{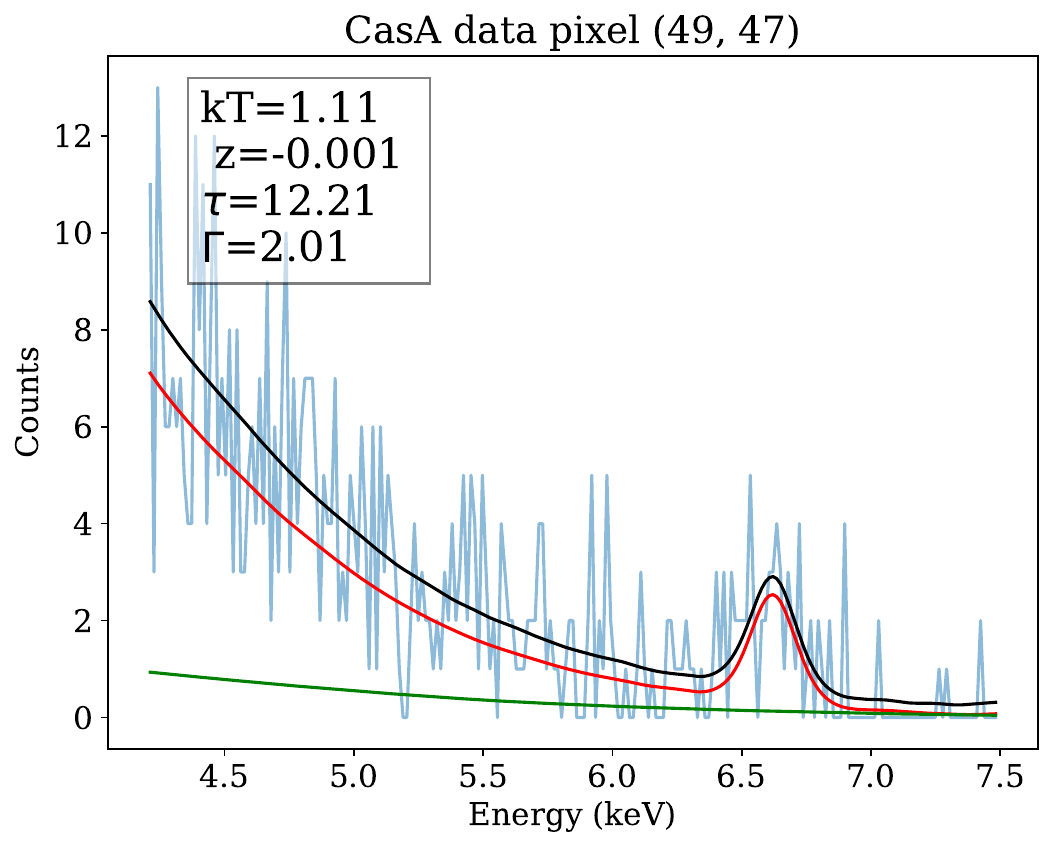}
\end{subfigure}
\\
\begin{subfigure}{\linewidth}
  \centering
    \includegraphics[width=\textwidth]{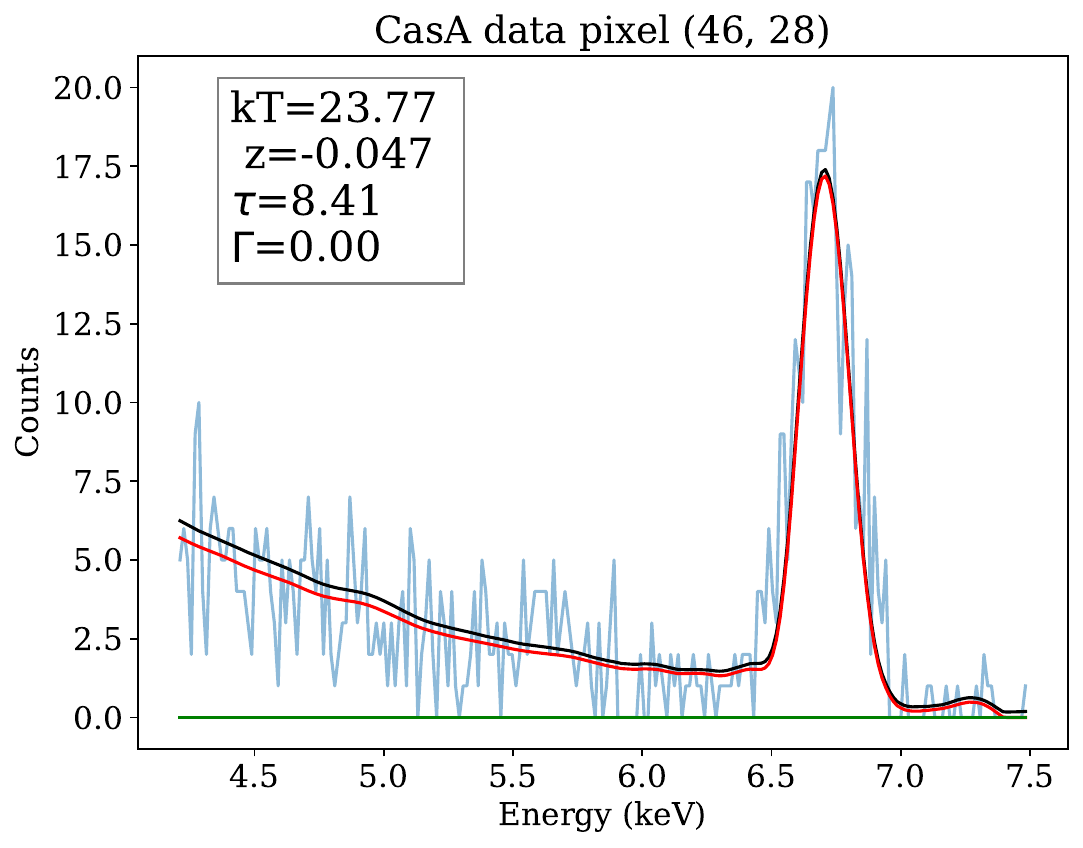}
\end{subfigure}
\\
\begin{subfigure}{\linewidth}
  \centering
    \includegraphics[width=\textwidth]{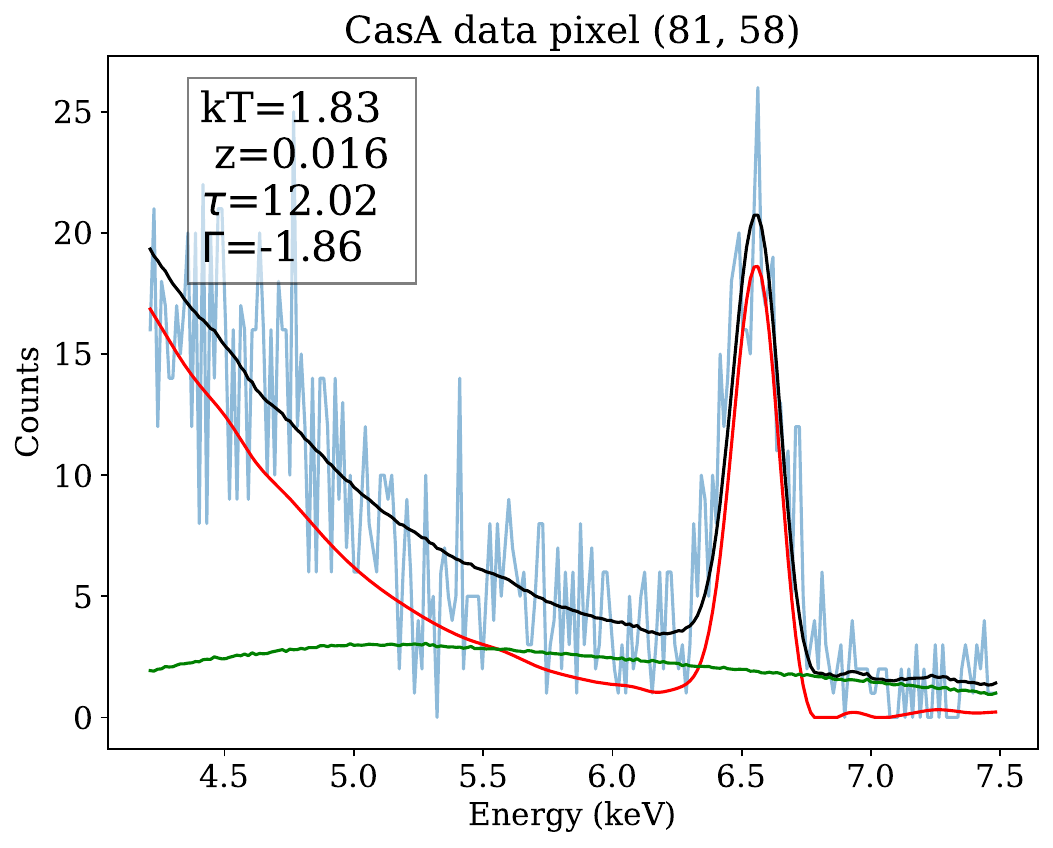}
\end{subfigure}
\caption{Example of three pixels fitted by SUSHI for the Cassiopeia data with their best-fit parameters. 
The temperature kT is in kiloelectronvolts (keV), and the ionization timescale $\tau$ is in log-scaled seconds per centimetre-cubed ($log_{10}$(cm$^{-3}$ s)).
The colors are as follows, blue: data; green: synchrotron component; red: thermal component; black: total fit.}
\label{fig:CasApixels}
\end{figure}

In Figure \ref{fig:CasApixels}, pixel (81,58), we observed that this may occur because the power law with a negative photon index can increase the amplitude of the iron line while keeping the continuum low, which is the effect of a high iron abundance. Pixel (46,28) shows that this issue is also likely to have caused the unrealistically high temperatures found in some pixels. However, of the fitted pixels, only 0.4$\%$ displayed a negative photon index, and 2$\%$ had temperatures above 4 keV, so this problem remains of limited impact. Most fitted pixels had no noticeable issue, such as pixel (49,47).

\subsection{Crab pulsar wind nebula}
\label{sec:crab}
Finally, we tested SUSHI in the simplest case, where there is only one component (no unmixing), applying it to data from the Crab pulsar wind nebula. In that case, only the synchrotron emission is present.
The data cube was constructed from the Chandra X-ray observation from the years 2000 and 2001 (ObsID: 1994, 1995, 1996, 1997, 1999, 2000, 2001). All the observations were stacked on a single cube with a spatial binning of 1''.

\begin{figure}
    \centering
    \includegraphics[width=\linewidth]{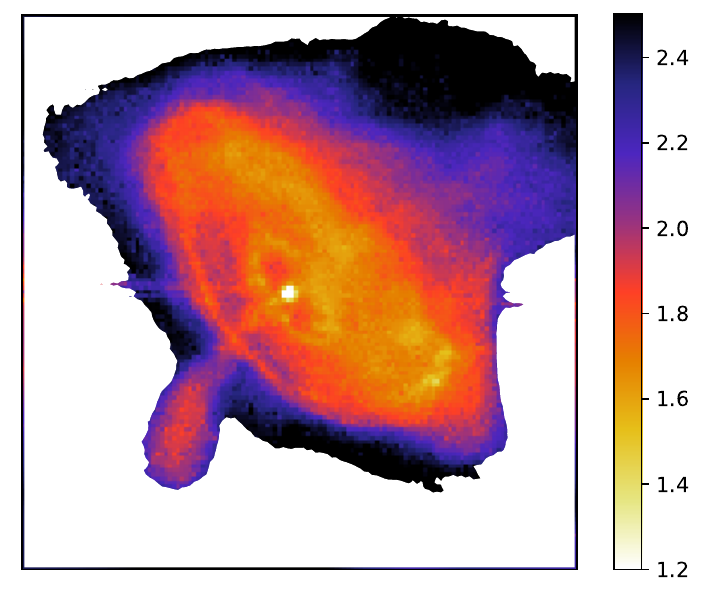}
    \caption{Photon index map of the Crab Nebula at a spatial resolution of 1". We note that the spectral index has not been corrected for the pile-up effect (see main text).}
    \label{fig:crab}
\end{figure}

The resulting photon index map from SUSHI is shown in Figure \ref{fig:crab}. It is consistent with past results, such as the index map in \cite{Mori_2004} (which used the same data set), but has a finer resolution thanks to the spatial regularization.

In this case, the IAE spectral model was trained on a collection of power laws with spectral indexes varying from 1.0 to 4.0 in the 0.8-7 keV energy band.
The absorption along the line of sight (modeled with $tbabs$ and the $wilm$ abundances table) was fixed to $N_{\rm H} = 0.44 \times 10^{22}$ cm$^{-2}$ \citep[see Table 1 of ][]{Weisskopf2004}.

We note that in the training set and in the fitted model, we assumed a single effective area file, while in reality the effective area (ARF file) changes slightly observation per observation and as a function of the pixel position in the camera.
We therefore tested the impact of this assumption by fitting the same spectrum but swapping the ARF file with that of another observation. The effect was minor on the reconstructed spectral index ($\Delta \Gamma \sim$0.01).
A similar test was performed by swapping the ARF files obtained at different camera positions, which also resulted in a minor effect ($\Delta \Gamma \sim$0.02).
Another effect not taken into account here is the pile-up effect. When more than one photon is recorded in a given pixel during a camera frame, the reconstructed energy is the sum of the photons' energies.
This effect produces an artificially harder photon index  (smaller index values) in regions of increased surface brightness.
For the brightest regions of the Crab Nebula, the shift in spectral index value is on the order of $\Delta \Gamma \sim$0.2 \citep[see Figure 2a of][]{Mori_2004}.

\subsection{Current limitations and way forward}
\label{sec:limits}
\paragraph{}
One current limitation of SUSHI is that it currently does not account for variations in the effective area across the image (it only takes one ARF file when training the model). However, it would not be difficult to generate a cube that accounts for the correction due to the varying effective area and to include that cube as a multiplicative factor when calculating the cost function to locally compensate for the variation of the effective area.

Further, the performance of SUSHI depends on the quality of the trained model, which can be limited in cases when the model is especially degenerate, is of high dimensionality, or has spectra displaying amplitude variations across many orders of magnitude. In addition to that, IAE models currently do not return physical parameters but latent parameters, and a classic least-square fit needs to be performed at the end to retrieve these parameters. All of these limitations are currently being investigated in order to apply SUSHI to increasingly complex data.

\section{Conclusions}
\paragraph{}
In this article, we have presented SUSHI (code available here\footnote{\url{https://github.com/jmlascar/SUSHI}}), an algorithm for non-stationary unmixing of hyperspectral images with spatial regularization of spectral parameters. Unlike most source separation methods, all physical components obtained by SUSHI vary in spectral shape and in amplitude across the hyperspectral image. For our spectral model, we used an IAE, and for our spatial regularization, we applied a sparsity constraint on the wavelet transform of the spectral parameter maps.

We applied SUSHI to a toy model meant to resemble the case study of supernova remnants in X-ray astrophysics. We compared the result to the one obtained by the classic method used in the literature (i.e., a 1D fit for each individual pixel). We find that SUSHI obtains better results, particularly when it comes to reconstructing physical parameters.

We then applied SUSHI to real data from the supernova remnant Cassiopeia A. The results obtained were realistic and in accordance with past findings, though we noticed that it was difficult for SUSHI to reconstruct the synchrotron component in patches where its amplitude was weak and the thermal component dominated, which we expected since the patches were large enough that the spatial regularization could not go around this difficulty.

Finally, we applied SUSHI to a simpler real case: X-ray Chandra data of the Crab Nebula, which only contains one component, the synchrotron emission. SUSHI was able to reconstruct a realistic synchrotron photon index reliably and with unprecedented detail.

Though we chose to study the case of X-ray astrophysics, SUSHI is also applicable at any other wavelength where hyperspectral images are observed. Thanks to the spatial regularization of the spectral parameters, it can unmix physical components with more accuracy than classic methods and 
map spectral properties even at fine scales.

\begin{acknowledgements}
The research leading to these results has received funding from the European Union’s Horizon 2020 Programme under the AHEAD2020 project (grant agreement n. 871158). This work was supported by CNES, focused on methodology for X-ray analysis. We thank S. Orlando for kindly providing the simulation from \cite{Orlando_2016}  to generate our toy model.
\end{acknowledgements}


\begin{appendix}
\section{Network architecture}
\label{sec:Appendix_IAE}
\def\arraystretch{1.4}
\begin{table*}
    \centering
    \begin{tabular}{|>{\centering\arraybackslash}m{0.1\textwidth}|>{\centering\arraybackslash}m{0.14\textwidth}|>{\centering\arraybackslash}m{0.14\textwidth}|>{\centering\arraybackslash}m{0.14\textwidth}|>{\centering\arraybackslash}m{0.14\textwidth}|>{\centering\arraybackslash}m{0.14\textwidth}|}
    \hline
         &\textbf{Thermal (toy model)}  &  \textbf{Thermal (Cassiopeia A data)} & \textbf{Synchrotron (toy model)}\ &\textbf{Synchrotron (Cassiopeia A data)}& \textbf{Synchrotron (Crab data)}\\
         \hline \hline
         Physical model & Equilibrium collisional ionized plasma emission (APEC)& Non-equilibrium collisional ionized plasma emission & \multicolumn{3}{c|}{Power Law}\\\hline
         Number of anchor points & 4 & 6 & 2 & 2 & 2 \\\hline
         Number of layers & 4 & 4 & 4 & 2 & 2 \\\hline
         Step size & $6\times10^{-4}$ &$4\times10^{-4}$ & $8\times10^{-4}$& $10^{-3}$&$10^{-3}$\\
         \hline
         Optimizer & \multicolumn{5}{c|}{Adaptive Gradient Algorithm (Adagrad)} \\
         \hline
         Activation function & \multicolumn{5}{c|}{Leaky Rectified Linear Activation (LReLU)}\\
         \hline
    \end{tabular}
    \caption{Information about the architecture of the five trained IAEs used in this work. }
    \label{tab:IAE_arch}
\end{table*}

\def\arraystretch{1.4}
\begin{table*}
    \centering
    \begin{tabular}{|>{\centering\arraybackslash}m{0.1\textwidth}|>{\centering\arraybackslash}m{0.17\textwidth}|>{\centering\arraybackslash}m{0.17\textwidth}|>{\centering\arraybackslash}m{0.12\textwidth}|>{\centering\arraybackslash}m{0.12\textwidth}|>{\centering\arraybackslash}m{0.12\textwidth}|}
    \hline
         &\textbf{Thermal (toy model)}  &  \textbf{Thermal (Cassiopeia A data)} & \textbf{Synchrotron (toy model)}\ &\textbf{Synchrotron (Cassiopeia A data)} & \textbf{Synchrotron (Crab data)}\\
         \hline \hline
         Physical model & Equilibrium collisional ionized plasma emission (APEC)& Non-equilibrium collisional ionized plasma emission & \multicolumn{3}{c|}{Power Law} \\\hline
         Energy range & $3-8.48$ keV & $4.2-7.4$ keV & $3-8.5$ keV & $4.2-7.4$ keV & $0.8-7$ keV \\\hline
         Variables & \makecell{kT $\in$ [0.8 keV, 6 keV] \\ z $\in$ [-0.033,0.033]} & \makecell{kT $\in$ [1 keV, 8 keV] \\ z $\in$ [-0.05,0.05] \\
         $\tau$ $\in$ [$10^{8}, 10^{12}$]} & $\Gamma$ $\in$ [1.5,3.5]& \multicolumn{2}{c|}{$\Gamma$ $\in$ [1,4]} \\\hline

    \end{tabular}
    \caption{Information about the training set of the five IAEs.}
    \label{tab:training_set}
\end{table*}
The IAE models trained in this work were coded in Python using the package JAX.\footnote{\url{https://jax.readthedocs.io}} We trained five networks: two to use on our toy model, two to use on real data from Cassiopeia A, and one to use on real data from the Crab Nebula. Details of the architecture of the five networks can be found in Table \ref{tab:IAE_arch}, and details about the training sets of each can be found in \ref{tab:training_set}.

\section{Model reconstruction quality}
To test the model's reconstruction quality, that is, whether the decoder generates spectra that look like the testing set, we forward-passed the testing set $\{s_i\}$ into the auto-encoder: $\hat{s}_i= D(I\odot E(s_i)$, where $I$ represents the projection of $E(s_i)$ onto the set of all affine combinations on the anchor points (see \cite{remi_thesis} section 2.3 for more details) and evaluated the residual between $s_i$ and $\hat{s}_i$: $R=(\hat{s}_i-s_i)/s_i$. If the IAE has been well trained, this value should be low.

\begin{figure}[H]
    \centering
    \includegraphics[width=0.8\linewidth]{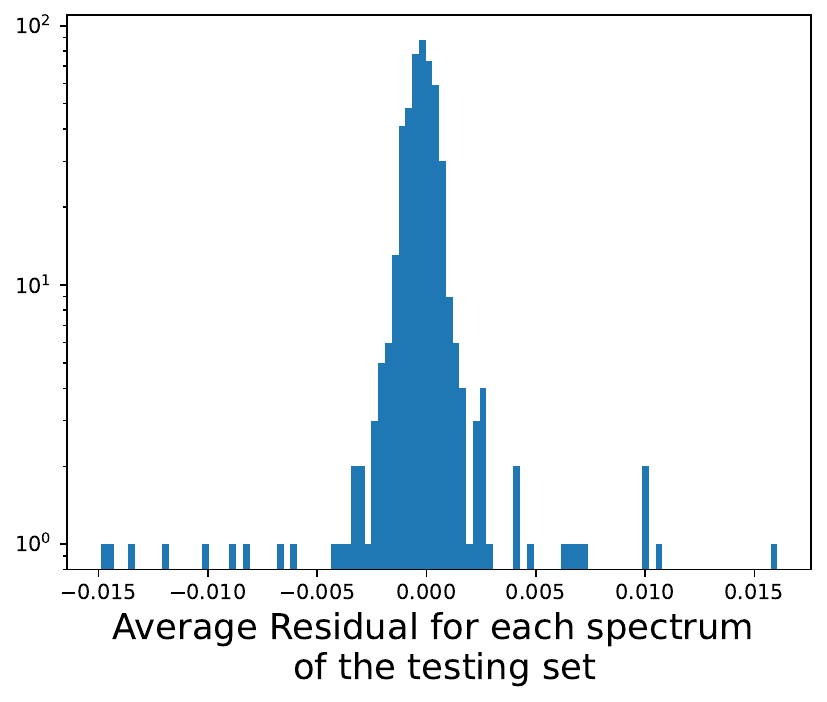}
    \caption{Histogram of the average residual $R=(\hat{s}_i-s_i)/s_i$ for the IAE trained for the thermal component of the toy model.}
    \label{fig:IAE_quality_thermaltoymodel}
\end{figure}

Figures \ref{fig:IAE_quality_thermaltoymodel} to \ref{fig:IAE_quality_synchCasA} show the distribution of that error for every member of the testing set of the five trained networks. Table \ref{tab:err_IAE} shows the average error for each of the networks.
\def\arraystretch{1.4}
\begin{table*}
    \centering
    \begin{tabular}{|>{\centering\arraybackslash}m{0.12\textwidth}|>{\centering\arraybackslash}m{0.16\textwidth}|>{\centering\arraybackslash}m{0.16\textwidth}|>{\centering\arraybackslash}m{0.12\textwidth}|>{\centering\arraybackslash}m{0.12\textwidth}|>{\centering\arraybackslash}m{0.12\textwidth}|}
    \hline
         &\textbf{Thermal (toy model)}  &  \textbf{Thermal (Cassiopeia A data)} & \textbf{Synchrotron (toy model)}\ &\textbf{Synchrotron (Cassiopeia A data)} & \textbf{Synchrotron (Crab data)}\\
         \hline \hline
         Reconstruction error  & -0.024$\%$ & 0.040$\%$ & 0.041$\%$ & -0.035$\%$ & -0.052$\%$
         \\
         \hline
    \end{tabular}
    \caption{Average reconstruction error between input test spectra and the spectra reconstructed by interpolating between anchor points}
    \label{tab:err_IAE}
\end{table*}

\begin{figure}[H]
    \centering
    \includegraphics[width=0.8\linewidth]{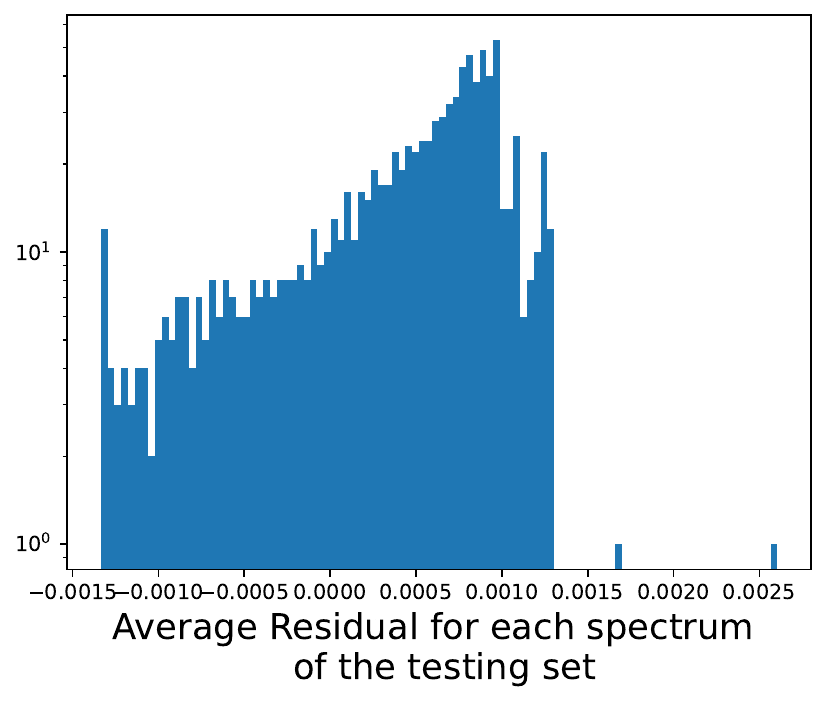}
    \caption{Histogram of the average residual $R=(\hat{s}_i-s_i)/s_i$ for the IAE trained for the synchrotron component of the toy model.}
    \label{fig:IAE_quality_synchtoymodel}
\end{figure}

\begin{figure}[H]
    \centering
    \includegraphics[width=0.8\linewidth]{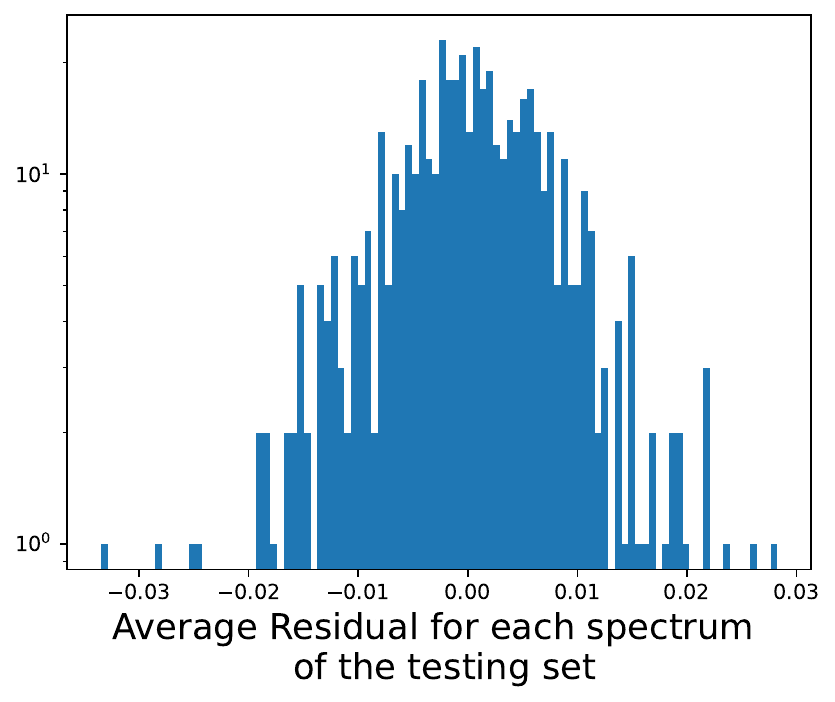}
    \caption{Histogram of the average residual $R=(\hat{s}_i-s_i)/s_i$ for the IAE trained for the thermal component of the Cassiopeia A data set.}
    \label{fig:IAE_quality_thermalCasA}
\end{figure}

\begin{figure}[H]
    \centering
    \includegraphics[width=0.8\linewidth]{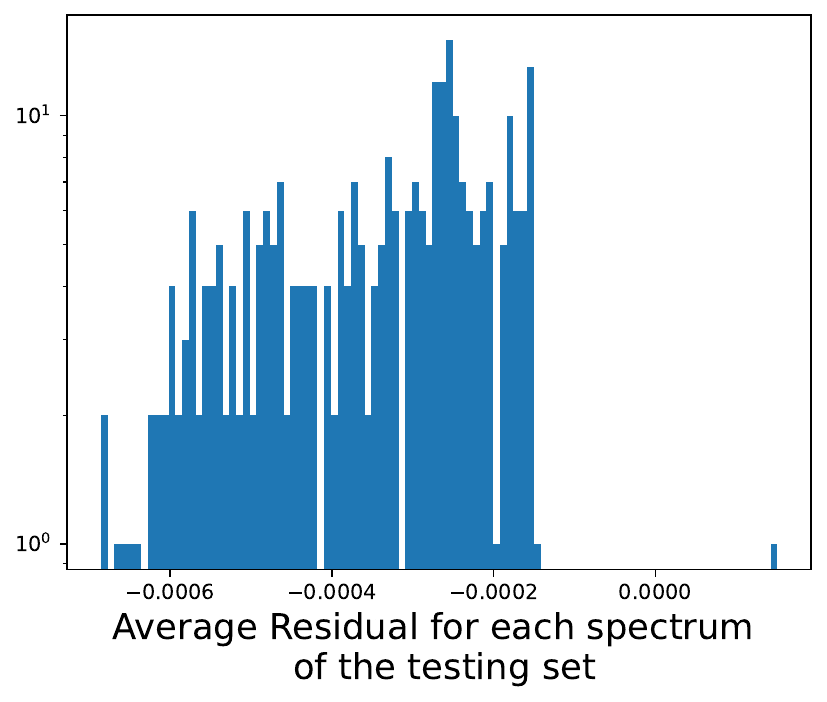}
    \caption{Histogram of the average residual $R=(\hat{s}_i-s_i)/s_i$ for the IAE trained for the synchrotron component c.}
    \label{fig:IAE_quality_synchCasA}
\end{figure}
\begin{figure}[H]
    \centering
    \includegraphics[width=0.8\linewidth]{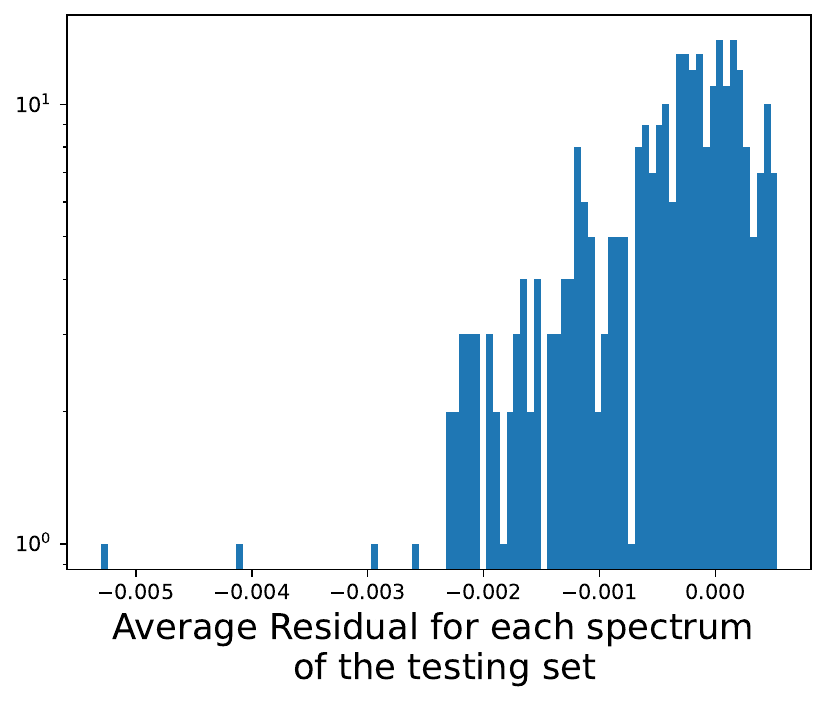}
    \caption{Histogram of the average residual $R=(\hat{s}_i-s_i)/s_i$ for the IAE trained for the synchrotron component of the Crab Nebula data set.}
    \label{fig:IAE_quality_synchCasA}
\end{figure}

\end{appendix}
\end{document}